\documentclass[a4paper,fleqn]{cas-dc}

\UseRawInputEncoding

\def\tsc#1{\csdef{#1}{\textsc{\lowercase{#1}}\xspace}}
\tsc{WGM}
\tsc{QE}
\usepackage[commandnameprefix=ifneeded]{changes}
\usepackage[authoryear]{natbib}

\usepackage{subfigure}
\usepackage{algorithm} 
\usepackage{algorithmic} 
\usepackage{multirow} 
\usepackage{amsmath}
\usepackage{xcolor}

\begin{document}
\let\WriteBookmarks\relax
\def\floatpagepagefraction{1}
\def\textpagefraction{.001}

\shorttitle{DySuse for Susceptibility Estimation}

\shortauthors{Y. Shi et~al.}

\title [mode = title]{DySuse: Susceptibility Estimation in Dynamic Social Networks}  

\author[1]{Yingdan Shi}[
]


\ead{ydshi98@stu.suda.edu.cn}

\credit{Conceptualization of this study, Methodology, Experiments, Writing - Original draft preparation}

\affiliation[1]{organization={Soochow University},
    city={Suzhou},
    postcode={215006}, 
    country={China}}

\author[1]{Jingya Zhou}[orcid=0000-0003-0721-7424]
\cormark[1]
\ead{jy_zhou@suda.edu.cn}
\credit{Funding acquisition, Conceptualization of this study, Supervision, Writing - review \& editing}

\author[1]{Congcong Zhang}[
]
\ead{cczhang_suda@outlook.com}
\credit{review \& editing}

\begin{abstract}
Influence estimation aims to predict the total influence spread in social networks and has received surged attention in recent years. 
{Most current studies focus on estimating the total number of influenced users in a social network,} and neglect susceptibility estimation that aims to predict the probability of each user being influenced from the individual perspective.
{As a more fine-grained estimation task, susceptibility estimation is full of attractiveness and practical value.
Based on the significance of susceptibility estimation and dynamic properties of social networks, we propose a task, called susceptibility estimation in dynamic social networks, which is even more realistic and valuable in real-world applications. Susceptibility estimation in dynamic networks has yet to be explored so far and is computationally intractable to naively adopt Monte Carlo simulation to obtain the results.
To this end, we propose a novel end-to-end framework DySuse based on dynamic graph embedding technology.} 
Specifically, we leverage a structural feature module to independently capture the structural information of influence diffusion on each single graph snapshot. 
Besides, {we propose the progressive mechanism according to the property of influence diffusion,} to couple the structural and temporal information during diffusion tightly. 
Moreover, a self-attention block {is designed to} further capture temporal dependency by flexibly weighting historical timestamps. 
Experimental results show that our framework is superior to the existing dynamic graph embedding models and has satisfactory prediction performance in multiple influence diffusion models.
\end{abstract}




\maketitle

\section{Introduction}
\label{introduction}
Social platforms generate tens of millions of messages every day. {As a fundamental task in the field of social computing, influence estimation focuses on mining such complex and rich information from the macro perspective to support many social applications such as viral marketing \citep{DBLP:journals/www/ZhouFWWL19}, social recommendation \citep{DBLP:conf/wsdm/ChenW21}, etc. Given a social network and an initial set of seed users, traditional influence estimation \citep{DBLP:journals/tst/wu20} aims at predicting how many users are influenced by the initial set of seed users (i.e., influence spread), while neglects individual susceptibility. Susceptibility estimation focuses on predicting the probability of an individual user being influenced from the microscopic perspective and has extensive applications in practice.}
{For instance, in marketing activities, enterprises can use susceptibility estimation to identify the potential users who are most likely to purchase their products in a social network. The probability of successful marketing for these target users is much higher than that of users with lower susceptibility. 
Therefore, some marketing interventions can be carried out on these target users, such as targeted recommendation advertisements and coupon incentives, to enable marketing activities more accurately. In addition to the application of marketing activities, there are other practical applications such as public opinion control and rumor propagation suppression.}
Xia et al. \citep{b17} propose a model named DeepIS for the first time to estimate the susceptibility. {But DeepIS is designed for static networks, while real-world social networks are dynamic and evolve over time, which significantly limits its application value. In this paper, we investigate susceptibility estimation in dynamic networks for the first time to deepen its practical significance and fill the gap in this field.}

Influence estimation has been studied along with influence maximization for years, and most work estimates the influence by repetitively simulating the influence diffusion process. To simulate the influence diffusion process, Independent Cascade (IC) diffusion model and Linear Threshold (LT) diffusion model have been proposed by Kempe et al. \citep{b11}. As two simple but fundamental diffusion models, the IC model assumes that a user may be influenced by one of its neighbors, and each influenced user has a certain probability of influencing its neighbors. In contrast, in the LT model, a user will be influenced once the total influence of all its neighbors exceeds a threshold. 
{Based on these two diffusion models, Kempe et al. \citep{b11} generalize them to a model called the triggering (TR) diffusion model. In the TR model, each node $v$ independently randomly selects a node set $T_v$ called trigger set from its in-neighbor set according to some distribution. Specifically, if a node $v$ has an in-neighbor who has been influenced at timestamp $t-1$ and belongs to its selected trigger set $T_v$, $v$ is influenced at timestamp $t$.}
Besides, many other diffusion models \citep{ICDM2010Yang,10.1145/988672.988739} have been proposed to extend these two models. Benefiting from the prosperity and development of these diffusion models, Monte Carlo (MC) simulation is the most widely used tool and method to estimate the influence in a specified diffusion model. 
Similarly, we can also employ MC simulation to estimate each user's susceptibility by simulating the influence diffusion process repeatedly and counting the frequency of each node being influenced under a sufficient number of simulations.
However, the MC simulation method is very time-consuming. Inspired by \citep{b17}, we intend to replace MC simulation with a deep learning-based regression task. Different from \citep{b17}, dynamic scenario requires us to capture both structural and temporal information to estimate susceptibility. Dynamic graph embedding is consequently a promising direction to explore for susceptibility estimation in dynamic social networks. However, influence diffusion is a dynamic propagation process coupled with dynamic network evolution, which makes the susceptibility estimation problem even more difficult. 
{Existing dynamic graph embedding models cannot be directly used to solve the dynamic susceptibility estimation problem due to the following challenges:}

\begin{itemize}
    \item \textbf{Incomplete modeling of social network dynamics}. The dynamic nature of social networks includes both topological evolution and feature evolution, where the former includes the addition or deletion of nodes and edges, and the latter considers changes in labels or features of nodes and edges. However, few existing dynamic graph embedding models take into account these evolution types at the same time, which prevents existing models from being applied to our task setting.

    \item \textbf{Lack of combination with the influence diffusion characteristics}. Network evolution is often modeled as a series of network snapshots, and then dynamic graph embedding is implemented by learning features across these snapshots.
    Influence diffusion is also conducted across these snapshots and has its own characteristics, i.e., the influence spread and susceptibility show a growth trend over time and have a dependency between adjacent snapshots.
    However, there is yet to be a solution to realize the dependency that conforms to the characteristics. 

\end{itemize}

{Facing the above challenges, 
we propose \textbf{DySuse}, a brand new framework originally tailored for the task of \textbf{sus}ceptibility \textbf{e}stimation in \textbf{dy}namic networks. Though some components of DySuse are inspired by prior work in the deep learning field, such as dynamic graph embedding and graph neural networks, we propose key methodologies in a novel framework to conquer the current challenges.}

To address \textbf{the first challenge}, DySuse considers all mentioned graph evolution types, especially the increase and decrease of nodes, which is challenging to handle by existing dynamic graph embedding models. 
For \textbf{the second challenge}, we propose a progressive mechanism that conforms to the characteristics of influence diffusion in dynamic networks so as to build the association between the initial features of different graph snapshots. Thus, DySuse can obtain the accumulated susceptibility of each node over time. {DySuse is not only conducive to applying research in this field to real-world application scenarios but also provides promising attempts to advance dynamic graph embedding.}

In short, we summarize the main contributions of this paper as follows:
\begin{itemize}
\item To the best of our knowledge, we are the first to explore susceptibility estimation in dynamic social networks.
\item We propose DySuse, a novel end-to-end framework, to enhance learning ability and improve estimation accuracy by taking full advantage of dynamic networks' topological and temporal information.
{\item The progressive mechanism we proposed is based on the monotonous growth property of influence diffusion in dynamic networks, and it tightly couples structural and temporal information captured by the different modules in DySuse and can reduce the estimation error effectively.}
{\item To model the network evolution, we propose a novel technology where we expand the network scale in advance, and regard isolated nodes as not in the network.} Consequently, DySuse can handle all mentioned types of graph evolution.
\item We conduct extensive experimental evaluations based on multiple datasets. The results demonstrate that DySuse consistently delivers superior performance compared to state-of-the-art baselines. 
\end{itemize}

The remainder of this paper is organized as follows. Section \ref{BB} summarizes the related work. Section \ref{CC} presents the preliminaries of the susceptibility estimation problem in dynamic networks. Section \ref{DD} elaborates on our model design. Section \ref{FF} conducts extensive experiments and presents the evaluation results as well as analyses. Section \ref{GG} concludes our work.

\section{Related work}
\label{BB}

\subsection{Influence Estimation}
Given a seed set, the previous studies on influence estimation focus on predicting the total influence spread. MC simulation is a straightforward method to estimate influence spread but is too time-consuming. To this end, considerable work has been carried out. Some work \citep{DBLP:conf/soda/BorgsBCL14,b13} utilize reverse reachable sets (RRS) for calculating the total influence spread. RRS is based on a novel sampling technique named Reverse Influence Sampling (RIS), which can be used in large-scale social networks. Recently, an increasing number of heuristic methods have been proposed. 
{Based on RIS technique, Guo et al. \citep{10.1145/3318464.3389740} propose an efficient random RR set generation algorithm that can improve the expected running time.}
Besides, Chen et al. \citep{b20} devise a new degree discount heuristics that can accelerate the calculation speed of marginal influence spread. 
Jung et al. \citep{b21} put forward a novel global influence ranking method IRIE that can use a small number of iterations to obtain the marginal influence spread.
{Guo et al. \citep{10.1145/3399661} propose a novel seeding strategy based on community structure and adopt sampling techniques to overcome the high complexity of continuous greedy.}
{Sun et al. \citep{10.1145/3336191.3371791} integrate the realistic scenario of spontaneous user adoption into influence diffusion and propose scalable algorithms which have approximate guarantee.}

{
Another series of works leverage deep learning techniques to study influence estimation.
Yang et al. \citep{b3} propose a novel multi-scale diffusion prediction model based on reinforcement learning and integrate the macroscopic diffusion size (influence spread) information into the RNN-based microscopic diffusion model.
Cao et al. \citep{b4} propose a novel model based on the GNN technique, called CoupledGNN, which uses two coupled graph neural networks to capture the interplay between influenced states and influence spread.
Wu et al.\citep{9835455} propose a novel framework HERI-GCN based on the GCN technique. Interestingly, HERI-GCN organically integrates RNN into the heterogeneous GCN, which can effectively improve the learning ability of the model.
Panagopoulos et al. \citep{b7} present a multi-task neural network to learn embeddings of nodes and predict influence spread. Though the multi-task model can capture the information of influence diffusion with relatively high quality, it is transductive and difficult to generalize to unseen data.} 
These works typically forecast the future influence spread of the specific content by mining the cascade information. Moreover, these preceding works are primarily devoted to predicting the total influence spread instead of the probability of a single user being influenced. 
To compensate for it, Xia et al. \citep{b17} propose a model named DeepIS that is tailored for susceptibility estimation, i.e., predicting the influenced probability of each user. However, most existing works, including DeepIS, ignore the dynamic nature of social networks in our cognitive domain. As a result, it is necessary to investigate susceptibility estimation in dynamic networks.

\subsection{Dynamic Graph Embedding}
Considering real-world networks' dynamic nature, recently, there has been a surge of interest in embedding dynamic graphs. The current works are mainly divided into the following categories:

\begin{itemize}

\item \textbf{Factorization approaches} use low-rank decompositions of time-dependent similarity measures to generate node embeddings over time. 
For example, \citep{b30} updates SVD decompositions incrementally, but it leads to error accumulation due to the perturbative approach. 

\item \textbf{Random walk-based approaches} generate not only sequences that capture topological dependencies, but also time-dependent contexts. 
For instance, Zhou et al. \citep{b41} use node2vec to train node embeddings at each timestamp and then adopt Orthogonal Procrustes to align node embeddings into a common space. It is trivial to see that generating random walks for each timestamp like \citep{b41} is time-consuming and costly.

\item \textbf{Deep learning approaches}
use deep learning technology to extract the topological properties of networks and capture temporal dependencies therein. Different architectures based on neural networks are adopted, i.e., RNNs, attention mechanisms \citep{10.5555/2969033.2969073}, CNNs \citep{NIPS2012_c399862d}. Deep learning approaches show remarkable performance in dynamic graph embedding and have broad research prospects. 

\end{itemize}



\section{Preliminaries}\label{CC}
We present the background of the susceptibility estimation problem and the related definitions. Here we take the IC model as an example for illustration purposes. Note that our work does not rely on the IC model; it is actually independent of diffusion models.

\begin{figure}[t]
\centering
\includegraphics[width=8cm]{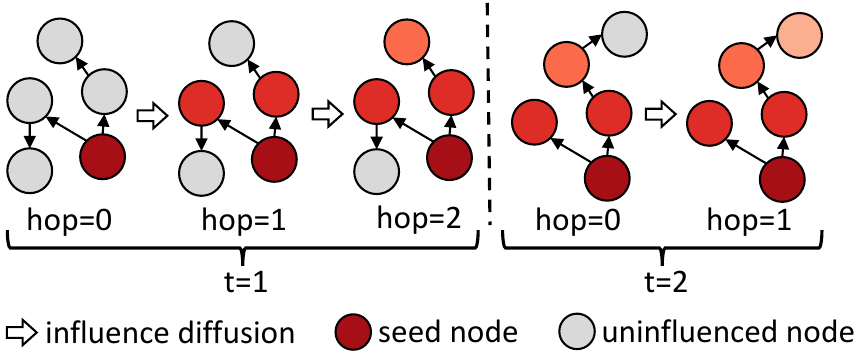}
\caption{An instance of influence diffusion in a dynamic graph.}
\label{dyinf}
\end{figure}

\textbf{Dynamic graph.} Given a sequence of the observed static graph snapshots, we have $G = \{G_1, G_2,. . . , G_T\}$, where $T$ is the number of timestamps. Each graph snapshot $G_t=(V_t,E_t)$ is characterized by a weighted adjacency matrix $\textbf{A}_t \in \mathbb{R}^{N \times N}$, where $N$ is the total number of nodes from all snapshots. To simulate real-world dynamic networks, we consider both topological evolution and feature evolution at the same time.

\textbf{Influence diffusion in a dynamic graph.} 
For a dynamic graph $G$, there is an initial node set as a seed set. In the beginning, the seed nodes in the seed set try to influence their neighbors. When the neighbors are influenced, they continue to influence their uninfluenced neighbors further. The process of influencing first-order neighbors is called 1-hop influence diffusion. Note that the hop count at each timestamp is unfixed but should not exceed an upper bound $hop_{max}$, where $hop_{max}$ is the maximum count of hops that the influence may diffuse across the entire graph. 
Based on the diffusion result on the last graph snapshot, the influence continues to diffuse at the next timestamp in the same way.
Figure~\ref{dyinf} shows an instance of influence diffusion in a dynamic graph, 
where the color saturation indicates the nodes influenced at different hops.

\textbf{Susceptibility estimation in a dynamic graph.}
Let $S$ be a seed set and $d$ be an influence diffusion instance in a dynamic graph $G$. We denote the state of node $v$ as $I_{S,d}(v)$. If node $v$ is influenced by the given seed set $S$ in the influence diffusion instance $d$, then $I_{S,d}(v) = 1$; otherwise $I_{S,d} (v) = 0$. Given a group of influence diffusion instances $\mathbb{D}$, the susceptibility of node $v$ is defined as $\Lambda_S(v) = \sum_d^{\mathbb{D}} I_{S,d}(v)/|\mathbb{D}|$. The problem is to estimate $\Lambda_S(v)$ for each node $v$ in $G$. In plain sight, the sum of susceptibility over all nodes equals the influence spread in $G$.

\begin{figure*}[t]
\centering

\includegraphics[width=0.9\textwidth]{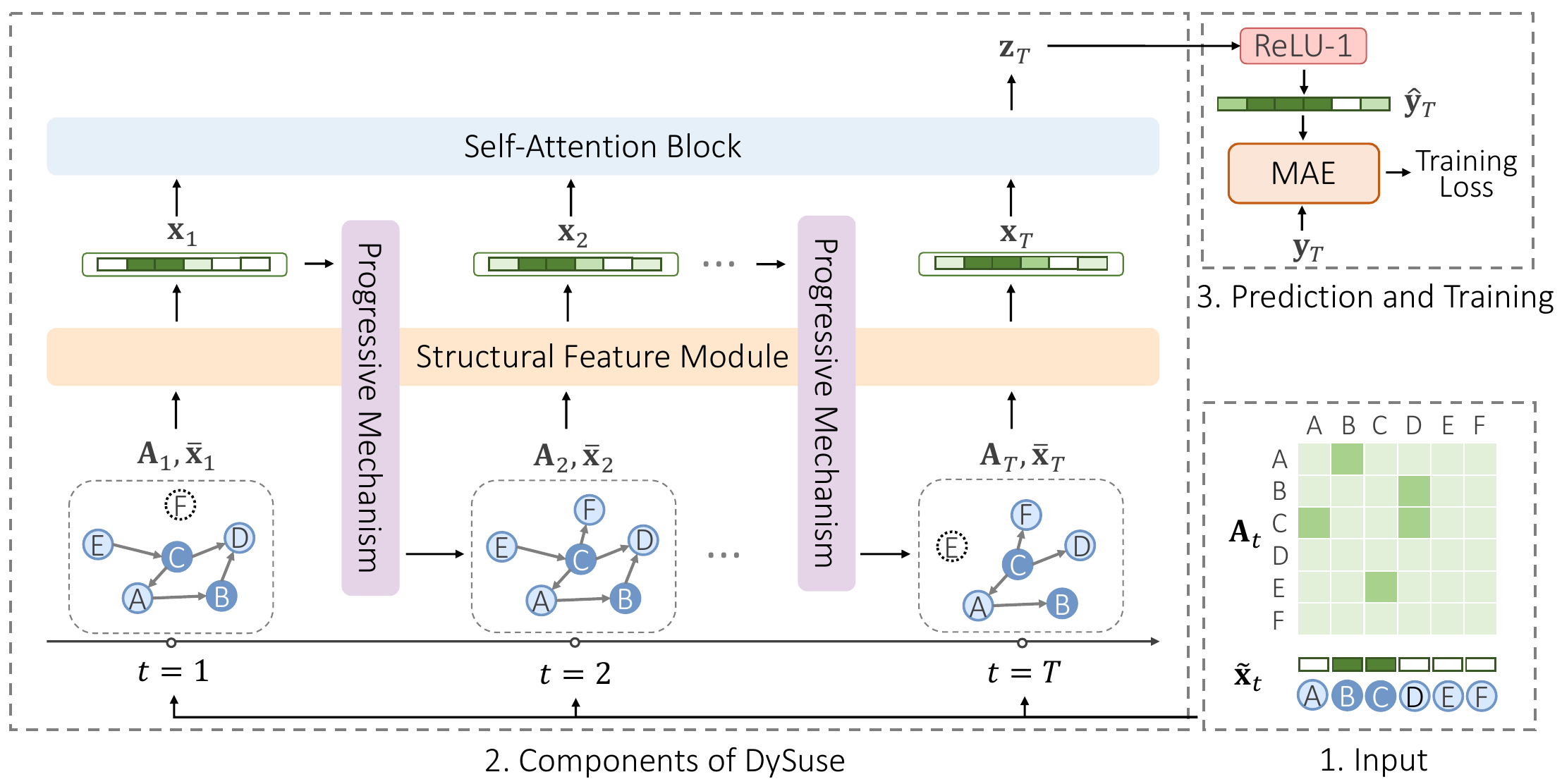}
\caption{{The DySuse framework for susceptibility estimation in dynamic social networks.}}
\label{dysuse}
\end{figure*}

\begin{figure*}[t]
\centering

\includegraphics[width=0.9\textwidth]{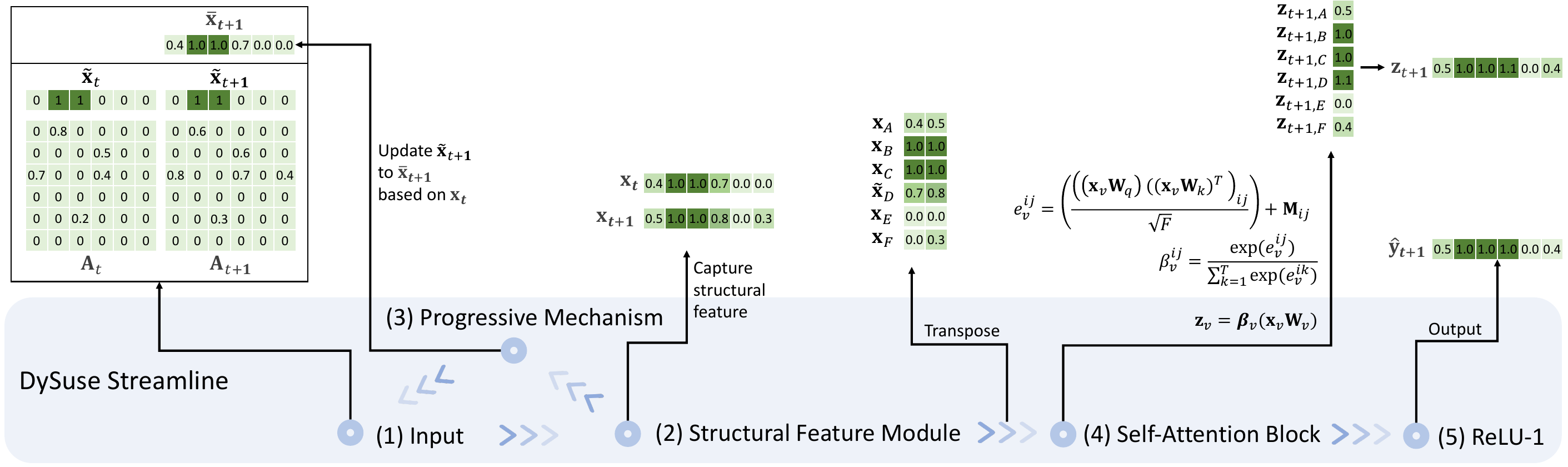}
\caption{{An example of DySuse used to predict. (1) Prepare inputs, including a weighted adjacency matrix and an initial feature vector for each graph snapshot. (2) Feed the inputs to the Structural Feature Module to capture the structural information. (3) Use the Progressive Mechanism to update the initial feature vector based on the output $\textbf{x}_t$ of the Structural Feature Module. (4) Transpose the feature vector and feed them into the Self-Attention Block to capture the temporal information. (5) Use $ReLU\raisebox{0mm}{-}1$ to limit the outputs between 0 and 1.}}
\label{dysuse_example}
\end{figure*}


\section{Our proposed method}

\label{DD}
We propose a novel framework called DySuse, whose overview and streamline example are illustrated in Figure~\ref{dysuse} {and Figure~\ref{dysuse_example}}. 
In DySuse, we construct a weighted adjacency matrix and generate an initial feature vector for each graph snapshot. 
Then, we design a structural feature module to capture the topological information of influence diffusion on each graph snapshot.
Meanwhile, we propose a progressive mechanism to tightly couple topological and temporal information by updating the initial feature vectors at the current timestamp with the output of the structural feature module at the previous timestamp. The progressive mechanism enhances the dependence between each graph snapshot's initial feature vector and enables DySuse to aggregate the susceptibility with time in a cumulative way.
After that, we feed the outputs of the progressive mechanism at each timestamp into the self-attention block to further capture the temporal dependence by flexibly weighting historical timestamps. 
Finally, we feed the output of the self-attention block to $ReLU\raisebox{0mm}{-}1$ to obtain the predicted susceptibility value at the last timestamp.

\begin{figure}[t]
\centering
\includegraphics[width=8cm]{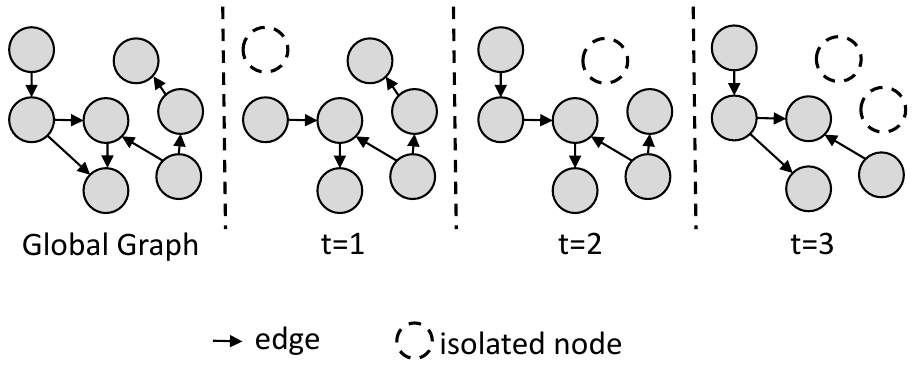}
\caption{Global graph and graph snapshots.}
\label{global}
\end{figure}

\subsection{Adjacency Matrix and Feature Construction} \label{input}

\textbf{The Weighted Adjacency Matrix.}
We classify the complex evolution in dynamic graphs into two types: topological evolution and feature evolution. To model both types, we construct a global graph that includes nodes and edges on all graph snapshots, and then we map the nodes in the global graph to each graph snapshot. As shown in Figure~\ref{global}, the dotted line nodes on each graph snapshot are isolated nodes mapped from the global graph. 
In fact, we do not need to be aware of the exact node number in advance, and we can even estimate the network growth rate based on historical changes and build a global graph.

After that, we can obtain the weighted adjacency matrix $\textbf{A}_t \in \mathbb{R}^{N \times N}$ for each $G_t$. Each edge $e_{t,(u,v)} \in E_t$ is associated with a weight $\textbf{A}_{t, (u,v)}$, representing the edge feature and indicating the probability of $v$ influenced by $u$ at timestamp $t$.

\textbf{The Initial Feature Vector.}
In each snapshot, only the susceptibility of seed nodes is 1, and the susceptibility of other nodes is initially set to 0. On the $t$-th graph snapshot, each node has a one-dimensional initial feature $\tilde{x}_{t,v}$ defined as
\begin{equation}
\tilde{x}_{t, v} = 
\begin{cases}
1, &   v \in S \ \ and \ \ v \in V_t, \\ 
0, &   otherwise.
\end{cases}
\end{equation}
Then, on graph snapshot $G_t$, we construct an initial feature vector $\tilde{\textbf{x}}_{t} = \{\tilde{x}_{t,1}, \tilde{x}_{t,2},..., \tilde{x}_{t,N}\}$, where $\tilde{\textbf{x}}_t \in \mathbb{R}^{N \times 1}$.

In summary, the weighted adjacency matrices reflect the network topology, probabilities of influence between nodes, and topology evolution of the dynamic network. The initial feature vectors contain the initial susceptibilities of nodes. Therefore, we can fully capture the susceptibility of each node by combining the weighted adjacency matrix and the initial feature vector at each timestamp.

\subsection{Structural Feature Module}\label{gnn}
A structural feature module is designed to predict the susceptibility of nodes on each graph snapshot. Multiple graph embedding models \citep{b54,b26,b4,b55} can be leveraged to implement this module, and we choose GNN to illustrate for simplicity.
GNN helps to generate the node feature vector by aggregating the topological information of each node and its surrounding nodes. 
{In the general paradigm, GNNs are designed to transform the features of each node in the graph and then aggregate information from the node's neighboring nodes to update the node's features. This process is repeated iteratively for every node in the graph.}
{Generally, given the weighted adjacency matrix $\textbf{A}_t$ and the initial feature vector $\tilde{\textbf{x}}_t$, the $l$-th layer of a GNN can be expressed as}
\begin{equation}
\begin{aligned}
\begin{gathered}
    h_{t,v}^{(l)} = AGGREGATE(\{ x_{t,u}^{(l)}, u \in N_t(v)\}), \\
    x_{t,v}^{(l+1)} = COMBINE(\{x_{t,v}^{(l)}, h_{t,v}^{(l)}\}),
\end{gathered}
\end{aligned}
\label{eq_agg_com}
\end{equation}
where $N_t{(v)}$ is the neighbors of node $v$ on the $t$-th graph snapshot, $h_{t,v}^{(l)}$ is the feature of node $v$ aggregated from its neighbors, and $x_{t,v}^{(l)}$ represents the node $v$'s feature at the $l$-th layer at timestamp $t$.
{According to the general paradigm of GNNs, the AGGREGATE($\cdot$) function is used to aggregate the features of the target node's neighbors, while the COMBINE($\cdot$) function is used to update the feature representation of the target node in the next layer using the aggregated representation from the AGGREGATE($\cdot$) function and the feature representation of the target itself.}
We define the input feature of node $v$ at timestamp $t$, i.e., $x^{(0)}_{t,v}$, as $\tilde{x}_{t,v}$.
Then we represent the structural feature module as
\begin{equation}
\label{equa-gnn}
x_{t,v}^{(l+1)} = 
\begin{cases}
GNN(\tilde{x}_{t,v}), &l = 0,\\ 
GNN(x_{t,v}^{(l)}), &l \geq 1.
\end{cases}
\end{equation}

Different graph snapshots are sequentially fed to the GNN with shared parameters. 
We mark the output of GNN at each timestamp as $\textbf{x}_t=\{x_{t,1}, x_{t,2}, ..., x_{t,N}\}$, which is a node feature vector and $\textbf{x}_t \in \mathbb{R}^{N \times 1}$.
Note that the different choices of the AGGREGATE($\cdot$) function and COMBINE($\cdot$) function form different GNNs. In other words, most GNNs, graph embedding models, and other susceptibility estimation methods in static networks can be adopted in our framework.

In this work, we find that the CoupledGNN \citep{b4} works best for the structural feature module. It is proposed to predict the total influence spread on the static graph, so we adapt it for our susceptibility estimation task.

As demonstrated in Figure~ \ref{coupledgnn}, CoupledGNN uses a one-dimensional value $\tilde{x}_{t,v}$ mentioned in Section\ref{input} to represent the susceptibility of node $v$ on the $t$-th graph snapshot, and the influence representation $\tilde{\textbf{r}}_{t,v}$ to express influence representation. The influence representation $\tilde{\textbf{r}}_{t,v}$ is initialized randomly in our work, while in the original CoupledGNN, it is initialized from node embedding obtained by DeepWalk \citep{10.1145/2623330.2623732}. Similar to $\tilde{\textbf{x}}_{t}$, $\tilde{\textbf{r}}_t = \{\tilde{\textbf{r}}_{t, 1}, \tilde{\textbf{r}}_{t,2}, ..., \tilde{\textbf{r}}_{t,N}\}$.

\begin{figure*}
\centering
\subfigure[State GNN.]{\includegraphics[width=0.45\textwidth]{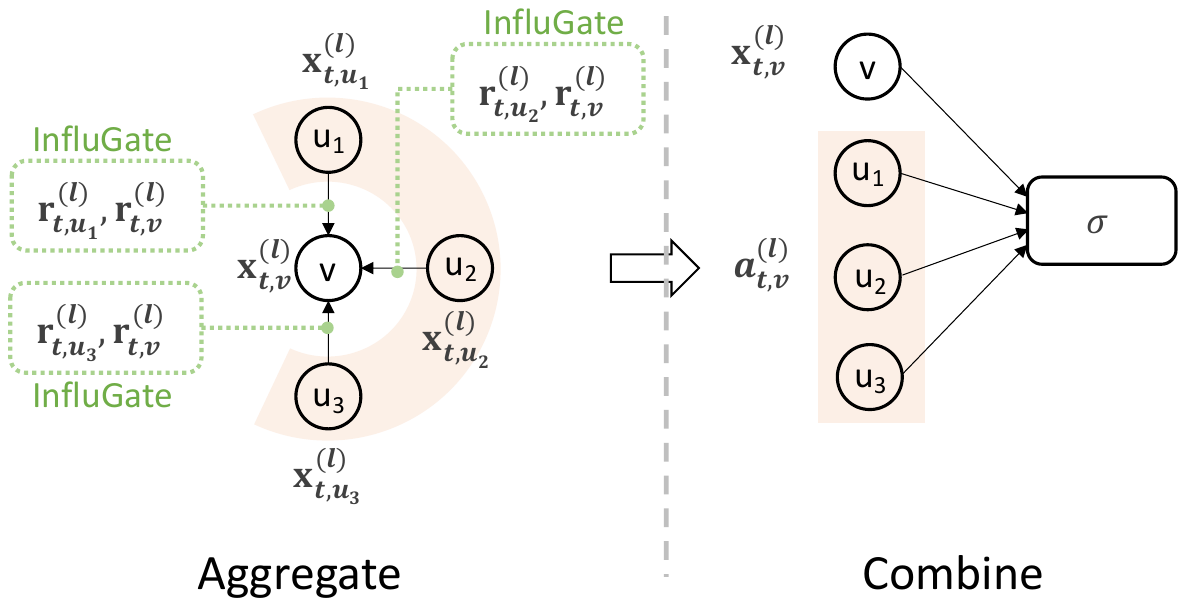}}
\hspace{0.5cm}
\subfigure[Influence GNN.]{\includegraphics[width=0.45\textwidth]{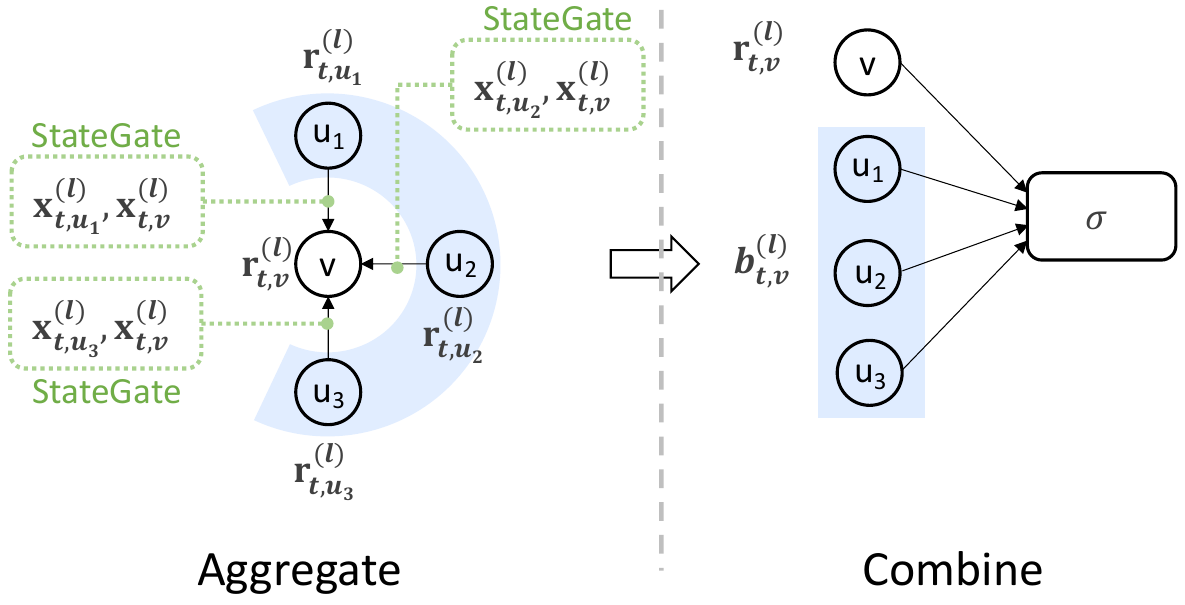}}

\caption{{Mechanisms of State GNN and Influence GNN.}}
\label{stategnn_influgnn}
\end{figure*}

CoupledGNN captures the interaction between the susceptibility and the total influence spread by using two graph neural networks named State GNN and Influence GNN. 
{The mechanisms of State GNN and Influence GNN are illustrated in Figure~\ref{stategnn_influgnn}}
The details of the two coupled graph neural networks are described as follows:

\begin{figure*}
\centering
\includegraphics[width=0.9\textwidth]{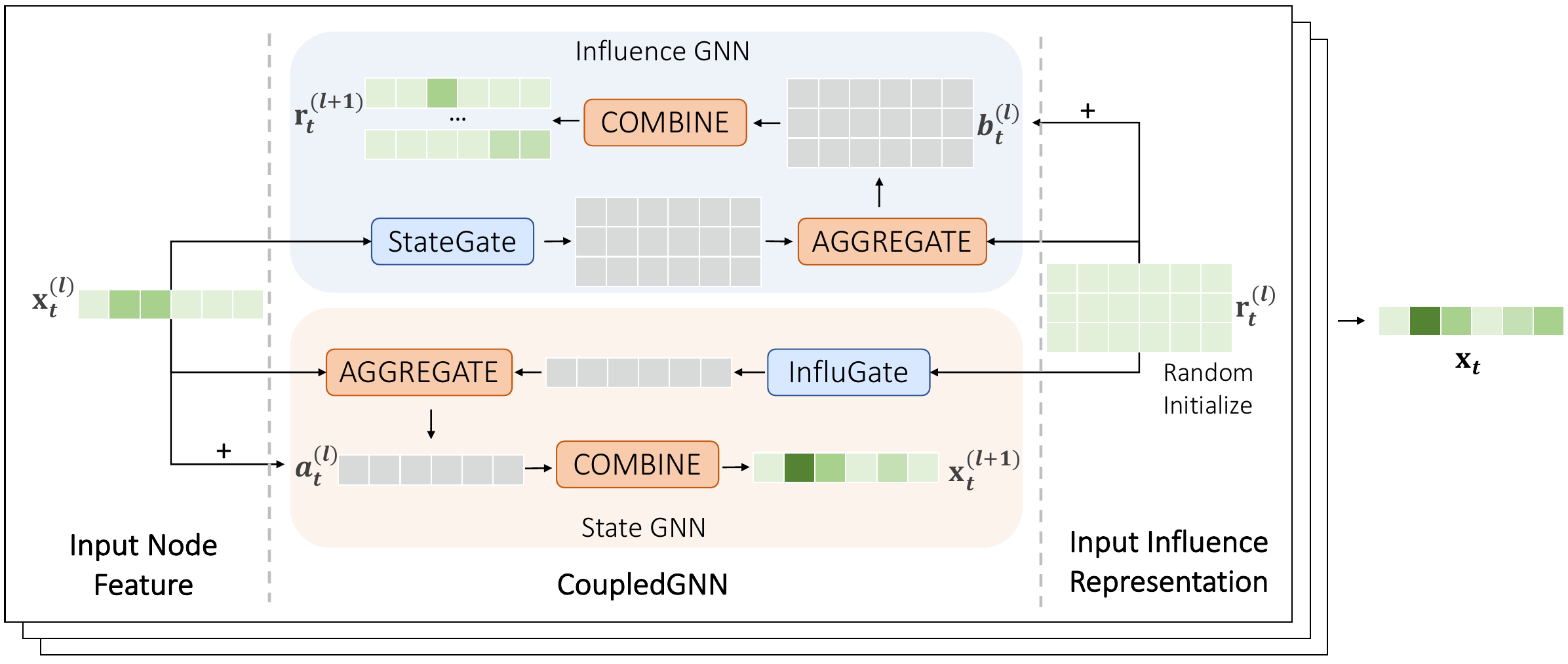}
\caption{{The framework of CoupledGNN.}}
\label{coupledgnn}
\end{figure*}

State GNN. It assigns the weight of influence between two nodes through an influence gating mechanism, i.e.,
\begin{equation}
InfluGate(\textbf{r}^{(l)}_{t,u} ,\textbf{r}^{(l)}_{t,v}) = \boldsymbol{\beta}^{(l)} [\textbf{W}^{(l)} \textbf{r}^{(l)}_{t,u} || \textbf{W}^{(l)} \textbf{r}^{(l)}_{t,v} ],
\end{equation}
where $\textbf{r}^{(l)}_{t,u} \in \mathbb{R}^{h^l}$ represents node $u$'s influence representation at the $l$-th layer at timestamp $t$, {$\textbf{W}^{(l)} \in \mathbb{R}^{h^{(l+1)} \times h^{(l)}}$ and $\mathbf{\beta}^{(l)} \in \mathbb{R}^{2h^{(l+1)}}$} are the weight matrix and weight vector respectively, {and $||$ denotes the vector concatenation operation}. Therefore, the expected influence aggregated from node $v$'s neighborhood can be formulated as follows,
\begin{equation}
a^{(l)}_{t,v} = \sum_{u \in N_t(v)} InfluGate(\textbf{r}_{t,u}^{(l)}, \textbf{r}_{t,v}^{(l)}) x^{(l)}_{t,u},
\end{equation}
{which is actually the AGGREGATE($\cdot$) function of State GNN.} Unlike vanilla CoupledGNN, we remove the self-activation parameter because our task aims at reflecting the influence diffusion by word-of-mouth, excluding the role of self-activation.
The susceptibility of node $v$ is updated to be the weighted sum of its neighborhood aggregation and the susceptibility of node $v$ at the last layer,
\begin{equation}
x^{(l+1)}_{t,v} = 
\begin{cases}
1, & i \in S, \\ 
\sigma(\mu^{(l)}_x x^{(l)}_{t,v} + \mu^{(l)}_a 
a^{(l)}_{t,v}), & i \notin S. \\ 
\end{cases}
\label{eq_combine1}
\end{equation}
$\sigma$ is the nonlinear activation function and $\mu^{(l)}_x$, {$\mu^{(l)}_a \in \mathbb{R}$} are both weight parameters. 
{Note that, Eq. \ref{eq_combine1} is actually a specific implementation of the COMBINE($\cdot$) function as mentioned in Eq. \ref{eq_agg_com}.}

Influence GNN. In this network, the neighborhood aggregation, {i.e., AGGREGATE($\cdot$) function,} is defined as
\begin{equation}
\begin{aligned}
\begin{gathered}
{StateGate(x_{t,u}^{(l)}) = MLPs(x_{t,u}^{(l)}),} \\
\textbf{b}_{t,v}^{(l)} = \sum_{u \in N_t(v)}StateGate(x_{t,u}^{(l)}) p_{t,uv} \textbf{W}^{(l)} \textbf{r}^{(l)}_{t,u},
\end{gathered}
\end{aligned}
\end{equation}
where $StateGate(\cdot)$ refers to the state gating, which is implemented by means of a 3-layer MLP. $\textbf{W}^{(l)} \in \mathbb{R}^{h^{(l+1)} \times h^{(l)}}$ is a weight matrix, and $p_{t,uv}$ is weight of edge $(u, v)$ on the $t$-th graph snapshot.
{Then, the COMBINE($\cdot$) function is used to update the influence representation and is defined as follows, }
\begin{equation}
 \textbf{r}^{(l+1)}_{t,v} =\sigma(\zeta_r^{(l)}\textbf{W}^{(l)}\textbf{r}^{(l)}_{t,v} +\zeta_b^{(l)}\textbf{b}_{t,v}^{(l)}),
\end{equation}
where $\zeta_r^{(l)}$, $\zeta_b^{(l)} \in \mathbb{R}$ are weight parameters.

We define the input node feature $x_{t,v}^{(0)}$ and input influence representation $r_{t,v}^{(0)}$ of node $v$ as, $x_{t,v}^{(0)} = \tilde{x}_{t,v}, \textbf{r}_{t,v}^{(0)} = \tilde{\textbf{r}}_{t,v}$. To sum up, we can represent the structural module using CoupledGNN as,
\begin{equation}
\label{equa-coupledgnn}
x_{t,v}^{(l+1)}, \textbf{r}_{t,v}^{(l+1)} = 
\begin{cases}
CoupledGNN(\tilde{x}_{t,v}, \tilde{\textbf{r}}_{t,v}), &l = 0,\\
CoupledGNN(x_{t,v}^{(l)}, \textbf{r}_{t,v}^{(l)}), &l \geq 1.
\end{cases}
\end{equation}

\subsection{Progressive Mechanism}\label{progressive}
The outputs of GNN contain the information of independent influence diffusion on each graph snapshot and do not contain temporal information. To this end, we integrate the progressive mechanism into the structural feature module to enhance the relevance and interactivity of structural information at different timestamps. 
The influence spread monotonously increases during the diffusion process, and susceptibility also has this property, which inspires us to compute the susceptibility over time cumulatively. In our progressive mechanism, we update the initial feature vector at timestamp $t$ with the output of GNN at timestamp $t-1$. 
Consequently, the structural information of each snapshot obtained from the GNN is based on the previous snapshot.
The update of $\tilde{x}_{t, v}$ through the progressive mechanism is denoted by 
\begin{equation}
\begin{aligned}
\overline{x}_{t, v} = 
\begin{cases}
x_{t-1, v}^{(L)}\ , & t \neq 0 \ \ and \ \ v \in V_{t-1},\\
\tilde{x}_{t,v}\ , & otherwise. \\
\end{cases}
\end{aligned}
\end{equation}
where $\tilde{x}_{t, v} \in \mathbb{R}$ is node $v$'s initial feature at timestamp $t$, and $x_{t-1, v}^{(L)}$ is the $L$-th layer's the output at timestamp $t-1$ for node $v$. Note that only the common nodes on both the $(t-1)$-th and $t$-th graph snapshots need to be updated. 
After being processed by the progressive mechanism, Eq.\ref{equa-gnn} is replaced by
\begin{equation}
\label{equa-gnn-2}
x_{t,v}^{(l+1)} = 
\begin{cases}
GNN(\overline{x}_{t,v}), &l = 0,\\
GNN(x_{t,v}^{(l)}), &l \geq 1.
\end{cases}
\end{equation}

\subsection{Self-attention Block}
To further capture temporal information, we incorporate a self-attention block in DySuse. From the output of the structural feature module, we obtain the node feature matrix on all snapshots, $\textbf{X} = \{\textbf{x}_1, \textbf{x}_2, ..., \textbf{x}_T\}$. By transposing $\textbf{X}$, we obtain the node feature vector on each graph snapshot $\textbf{x}_v=\{x_{1,v}, x_{2,v}, ..., x_{T,v}\}$, and $\textbf{x}_v \in \mathbb{R}^{T \times 1}$. Besides, we add position embeddings $\{p_1, p_2, ..., p_T\}$ by encoding the absolute temporal position of each timestamp to capture the sequential information. 
Then, the input of this block is updated by $\textbf{x}_v = \{x_{1,v}+p_1, x_{2,v}+p_2, ..., x_{T,v}+p_T\}$.

The output of the self-attention block is a temporal sequence of susceptibility for each node $v$, represented by $\textbf{z}_v = \{z_{1,v}, z_{2,v}, ..., z_{T,v}\}$, and $\textbf{z}_v \in \mathbb{R}^{T \times 1}$. We leverage the Scaled Dot-Product Attention \citep{b51} to learn the susceptibility of each node on different graph snapshots. We map the input $\textbf{x}_v$ into different feature spaces through linear projection matrices $\textbf{W}_q$, $\textbf{W}_k$ and $\textbf{W}_v \in \mathbb{R}^{D \times F}$, where $D$ ($F$) is the input (output) dimension and here $D=F=1$. Specifically, we have $\textbf{Q}_v = \textbf{x}_v\textbf{W}_q$ for queries, $\textbf{K}_v = \textbf{x}_v\textbf{W}_k$ for keys, and $\textbf{V}_v = \textbf{x}_v\textbf{W}_v$ for values. The self-attention layer is defined by
\begin{equation}
\begin{aligned}
\begin{gathered}
\textbf{z}_v = \boldsymbol{\beta}_v(\textbf{x}_v\textbf{W}_v),\\
\beta_v^{ij} = \frac{exp(e_v^{ij})}{\sum^T_{k=1}exp(e_v^{ik})}, \\
e_v^{ij} = (\frac{((\textbf{x}_v\textbf{W}_q)(\textbf{x}_v\textbf{W}_k)^T)_{ij}}{\sqrt{F}})+M_{ij},
\end{gathered}
\end{aligned}
\end{equation}
where $\beta_{ij}^v$ is the attention weight between timestamps $i$ and $j$ for node $v$, {$\boldsymbol{\beta}_v \in \mathbb{R}^{T \times T}$ is the attention weight matrix obtained by the multiplicative attention function.} Considering that the influence diffusion is unidirectional in time, we adopt a mask matrix {$\textbf{M} \in \mathbb{R}^{T \times T}$}, where $M_{ij}$ is defined as follows, 
\begin{equation}
    M_{ij}=
    \begin{cases}
    0, &   i \leq j, \\
    -\infty, &   otherwise.
    \end{cases}
\end{equation}
If $M_{ij} = - \infty$, it means the attention from timestamp $i$ to $j$ is off and $\beta_v^{ij}=0$.

\subsection{Prediction and Training}\label{PT} 

Given the weighted adjacency matrices and the initial feature vectors, we are supposed to estimate the susceptibility for each node at the last timestamp. The susceptibility of node $v$ at timestamp $T$ is calculated by
\begin{equation}
    \hat{y}_{T,v} = ReLU\raisebox{0mm}{-}1 (z_{T,v}),
\end{equation}
where $z_{T,v}$ is the output of the self-attention block at timestamp $T$ for node $v$. Inspired by \citep{b53}, we cap the $ReLU$ units at 1 to limit the results between 0 and 1, i.e., $ReLU\raisebox{0mm}{-}1(x)=\min(\max(x,0), 1)$. 
The eventual objective is to minimize the following loss function, i.e., Mean Absolute Error (MAE), denoted as,
\begin{equation}
L(\Theta) = \sum_{i=1}^{C} \sum_{v=1}^{N} |y_{T,v}-\hat{y}_{T,v}|,
\end{equation}
where $y_{T,v}$ is the corresponding ground truth, $C$ is the total number of training seed sets, and $\Theta$ denotes all parameters in DySuse.

\subsection{{Complexity Analysis}}\label{complexity_analysis} 
{The complexity of DySuse mainly depends on three parts: the Structural Feature Module, the Progressive Mechanism, and the Self-Attention Block. For the Structural Feature Module, its complexity depends on its specific implementation model. Here we take CoupledGNN as an example, for the dynamic graph with $N$ nodes, $M$ edges, and $T$ graph snapshots, the complexity of the State GNN is $O(T (Nh'h+Mh))$ where $h'$ and $h$ are the input and output dimensions, respectively, and the complexity of Influence GNN is $O(T (N+Nh'h+Mh))$.
Thus, the complexity of CoupledGNN in DySuse with a single layer is $O(T (Nh'h+Mh))$.}
{For the Progressive Mechanism, it takes $O(T N)$ to update the initial feature vectors.}
{In the Self-Attention Block, the complexity is $O(T^2 N)$ for a single forward propagation.}
{To sum up, the complexity of DySuse with a $1$-layer CoupledGNN as Structure Feature Module and a $1$-layer Self-Attention Block is $Max(O(T (Nh'h+Mh)), O(T^2 N))$.}

\section{Experiments}\label{FF}
\subsection{Experimental Settings}
\textbf{Datasets.} We conduct extensive experiments on one synthetic dataset and four real-world datasets available at \url{http://snap.stanford.edu/data} and \url{http://konect.cc/networks/}. The details of these datasets are shown in Table~\ref{tab1}. 

$\bullet$ BA\citep{b46}. We generate a static synthetic graph by using the Barab\'{a}si-Albert (BA) growth model, which follows a power-law distribution.

$\bullet$ Epinions\citep{b47}. It is a who-trust-whom social network. The nodes represent the users in the Epinions, and directed edges denote the trust relationship between user nodes. Epinions is a dynamic network with only two evolution types: node additions and edge additions.

$\bullet$ Enron\citep{b48}. It is an Enron email communication network related to email communication, where nodes represent employees and edges annotated with timestamps represent email communications. Similarly, the dataset has only two evolution types like Epinions.

$\bullet$ Facebook\citep{b49}. It includes a part of friend lists abstracted from the Facebook social network. It is a static network, so we randomly add and delete some nodes and edges at each timestamp to make it possess dynamic characteristics. The processing details will be presented later.

$\bullet$ Digg\citep{b50}. It is a reply network collected from the Digg news website, where nodes represent the website users and directed edges represent their reply behaviors. The evolution types of this dataset only include node and edge additions.


\begin{table}[t]
\caption{Dataset statistics}
\scriptsize
\begin{center}
\begin{tabular}{c c c c c c}
\hline
Dataset& Category& Type& \#Node& \#Edge& \#T  \\
\hline
BA& Synthetic& Undirected& 0.1K& 0.3K& 5  \\
Epinions& Social& Undirected& 1.2K& 5.0K& 5  \\
Enron& Communication& Undirected& 1.9K& 2.3M& 5  \\
Facebook& Social& Directed& 4.0K& 8.8M& 10  \\
Digg& Communication& Undirected& 1.1M& 4.2M& 10  \\
\hline
\end{tabular}
\label{tab1}
\end{center}
\end{table}

\textbf{Processing of Datasets.}
Considering that the existing datasets do not contain the traces of all evolution types of the dynamic network as we mentioned before, we further process the above datasets as follows: For datasets Epinions, Enron, and Digg, the edges in these datasets have timestamp information, and they are sorted in ascending temporal order in each dataset. For datasets BA and Facebook, there is no temporal information; we annotate the timestamps for edges according to the storage order of these edges within the original datasets and sort these edges as mentioned above.

\begin{itemize}
    
\item When timestamp $t=1$, 
we extract some of the edges in chronological order to construct a subgraph for each dataset, and select the largest connected component from the subgraph as the initial snapshot $G_1$, while the remaining edges will be added to the subsequent snapshots.

\item When timestamp $t>1$, 
\emph{for each real-world dataset},
i) we randomly delete 0$\sim$2\textperthousand{} of nodes and 0$\sim$2\textperthousand{} of edges at each timestamp; 
ii) to add nodes, we randomly generate 0$\sim$2\textperthousand{} nodes at each timestamp and let them randomly connect to some nodes in the previous graph snapshot; 
iii) to add edges, we add 0$\sim$2\textperthousand{} of nodes to the previous graph snapshot as the current graph snapshot. 
\emph{For the synthetic dataset}, the fraction of additions and deletions of nodes is set to 0$\sim$5$\%$, while the fraction of additions and deletions of edges is set to 0$\sim$10$\%$. As a result, the synthetic dataset, i.e., BA, is more dynamic than real-world datasets.
\end{itemize}

We set $T=10$ timestamps for those relatively large datasets, i.e., Facebook and Digg, and set $T=5$ timestamps for other datasets.
By following the related work \citep{b13}, the weight of edge $(u,v)$ is set as 
$w_{u,v} = \frac{1}{d_{in}(v)}$ where $d_{in}(v)$ is the in-degree of node $v$.
The weights will change along with the addition or deletion of edges, indicating that the features of edges evolve across different timestamps.
 Moreover, the seed size $k$ is chosen from $\{50, 100, 150, 200, 250\}$ for these real-world datasets and $\{5, 10, 15, 20, 25\}$ for the synthetic dataset BA. For each $k$, we generate 99 random seed sets and 1 seed set with top-$k$ node degrees. 

\textbf{Ground truth.}
The ground truth generation is similar to the existing works like \citep{10.1145/3084457,b17}. According to the influence diffusion in dynamic networks, we run 1000 MC simulations in each dataset to obtain the average probability of each node being influenced as the ground truth.
We use ground truth in different diffusion models to train and evaluate our framework.

\textbf{Baselines.}\label{Baselines}
We compare DySuse with the state-of-the-art baselines on dynamic graph embedding and adapt them to accommodate all types of graph evolution:

\emph{EvolveGCN} \citep{b34}. Instead of directly adopting a recurrent layer to embed the temporal information, EvolveGCN adopts RNN to train the weights of GCN.

\emph{DySAT} \citep{b32}. It learns node representations via joint self-attention along two dimensions, i.e., the temporal dynamics and the structural neighborhood.

\emph{TSGNet} \citep{b31}. It is proposed to exploit interaction links for node classification. It captures complex interaction patterns by combining GNNs and recurrent networks.

\begin{table*}[t]
\caption{The MAE value of different methods for susceptibility estimation on all datasets in the IC model. For all experimental data, we retain 3 digits after the decimal point.}
\label{baseline}
\resizebox{2\columnwidth}{!}{
\begin{tabular}{p{1.3cm}| p{1.4cm}<{\centering}| p{1.5cm}<{\centering} p{1.4cm}<{\centering} p{1.4cm}<{\centering} p{1.4cm}<{\centering} p{1.4cm}<{\centering} p{1.4cm}<{\centering} p{1.4cm}<{\centering} p{1.4cm}<{\centering} }
\hline
\multicolumn{1}{l}{Dataset} & \begin{tabular}[c]{@{}c@{}}Seed Size\end{tabular} & EvolveGCN & DySAT & TSGNet & \begin{tabular}[c]{@{}c@{}}DySuseG\end{tabular} & \begin{tabular}[c]{@{}c@{}}DySuseS\end{tabular} & \begin{tabular}[c]{@{}c@{}}DySuseM\end{tabular} & \begin{tabular}[c]{@{}c@{}}DySuseP\end{tabular} & \begin{tabular}[c]{@{}c@{}}DySuseC\end{tabular} \\
\hline
\hline

\multirow{5}{*}{BA} & 5 & 0.238 &0.162 & 0.228 & 0.233 & 0.193 & 0.346 & 0.131 & \textbf{0.090} \\
& 10 & 0.259 & 0.284 & 0.250 & 0.249 & 0.185 & 0.256 & 0.108 & \textbf{0.080} \\
& 15 & 0.324 & 0.278 & 0.249 & 0.244 & 0.152 & 0.249 & 0.139 & \textbf{0.100} \\
 & 20 & 0.404 & 0.216 & 0.225 & 0.211 & 0.181 & 0.145 & 0.102 & \textbf{0.059} \\
& 25 & 0.390 & 0.227 & 0.242 & 0.236 & 0.238 & 0.234 & \textbf{0.065} & 0.067 \\
\hline

\multirow{5}{*}{Epinions} 
& 50 & 0.450 & 0.367 & 0.382 & 0.260 & 0.225 & 0.263 & 0.086 & \textbf{0.085} \\
& 100 & 0.485 & 0.374 & 0.399 & 0.293 & 0.212 & 0.212 & 0.115 & \textbf{0.102} \\
& 150 & 0.582 & 0.374 & 0.385 & 0.289 & 0.240 & 0.211 & 0.106 & \textbf{0.091} \\
& 200 & 0.591 & 0.361 & 0.373 & 0.287 & 0.264 & 0.201 & 0.101 & \textbf{0.099} \\
& 250 & 0.436 & 0.310 & 0.366 & 0.291 & 0.232 & 0.233 & 0.098 & \textbf{0.075} \\

\hline

\multirow{5}{*}{Enron} 
& 50 & 0.327 & 0.327 & 0.375 & 0.297 & 0.232 & 0.154 & 0.106 & \textbf{0.082} \\
& 100 & 0.326 & 0.346 & 0.401 & 0.288 & 0.209 & 0.190 & 0.106 & \textbf{0.078} \\
& 150 & 0.448 & 0.336 & 0.444 & 0.286 & 0.233 & 0.183 & 0.093 & \textbf{0.073} \\
& 200 & 0.518 & 0.334 & 0.428 & 0.299 & 0.211 & 0.168 & 0.094 & \textbf{0.074} \\
& 250 & 0.527 & 0.317 & 0.422 & 0.308 & 0.241 & 0.208 & 0.083 & \textbf{0.073} \\

\hline

\multirow{5}{*}{Facebook}
& 50 & 0.617 & 0.270 & 0.355 & 0.268 & 0.225 & 0.191 & 0.103 & \textbf{0.005} \\
& 100 & 0.483 & 0.264 & 0.351 & 0.255 & 0.217 & 0.146 & 0.080 & \textbf{0.033} \\
& 150 & 0.589 & 0.251 & 0.254 & 0.239 & 0.244 & 0.231 & 0.076 & \textbf{0.028} \\
& 200 & 0.390 & 0.226 & 0.246 & 0.232 & 0.208 & 0.227 & 0.071 & \textbf{0.022} \\
& 250 & 0.603 & 0.216 & 0.296 & 0.216 & 0.202 & 0.233 & 0.072 & \textbf{0.029} \\

\hline

\multirow{5}{*}{Digg}
& 50 & 0.555 & 0.285 & 0.294 & 0.284 & 0.277 & 0.231 & 0.070 & \textbf{0.057} \\
& 100 & 0.445 & 0.299 & 0.331 & 0.271 & 0.240 & 0.219 & \textbf{0.057} & 0.066 \\
& 150 & 0.491 & 0.283 & 0.307 & 0.279 & 0.264 & 0.189 & 0.063 & \textbf{0.052} \\
& 200 & 0.532 & 0.321 & 0.348 & 0.308 & 0.284 & 0.213 & 0.067 & \textbf{0.049} \\
& 250 & 0.560 & 0.301 & 0.342 & 0.301 & 0.254 & 0.260 & 0.070 & \textbf{0.050} \\
\hline
\end{tabular}}
\end{table*}

For our proposed DySuse framework, five variants of DySuse are derived. Specifically, we implement the structural feature module by adopting GCN, GraphSAGE, MONSTOR, DeepIS, and CoupledGNN.

\emph{DySuse-GCN} \citep{b26} (DySuseG). 
GCN generates a node's representation by aggregating its own feature and neighbors' features. 

\emph{DySuse-GraphSAGE} \citep{b54} (DySuseS). 
GraphSAGE is an inductive model that generates embeddings for nodes by sampling and aggregating their neighborhood features.

\emph{DySuse-MONSTOR} \citep{b55} (DySuseM). 
MONSTOR originally predicts the total influence spread by stacking multiple GCN models, and we adjust it to predict the susceptibility of each node. The ground truth generation is similar to existing works like \citep{10.1145/3084457,b17}. According to the influence diffusion in dynamic networks, we run 1000 MC simulations in each dataset to obtain the average probability of each node being influenced as the ground truth.
We use ground truth in different diffusion models to train and evaluate our framework.

\emph{DySuse-Propagation Scheme} \citep{b17} (DySuseP). 
The propagation scheme is the key component of DeepIS \citep{b17}, which considers the characteristics of influence in different diffusion models. 
We select the component instead of DeepIS to implement this baseline since DeepIS even cannot converge on our task when the iterative value increases up to 5, while the propagation scheme does not have this problem.

\emph{DySuse-CoupledGNN} \citep{b4} (DySuseC). CoupledGNN devises two graph neural networks that are coupled with each other to effectively capture the interplay between the susceptibility and the influence spread. CoupledGNN is originally used to predict cascade popularity, and we alter it to predict the susceptibility of nodes in static networks.

\textbf{Implementation Details.}
We elaborate on the implementation details from the following two aspects:

\emph{1)} Dynamic graph embedding baselines. For EvolveGCN, we adopt a 2-layer structure and apply an MLP to support the susceptibility estimation tasks, {as reported in their original papers \citep{b34}}. For DySAT, we use its default parameters in the publicly available implementation. For TSGNet, we adopt a 3-layer structure with 64 hidden units. 

\emph{2)} Variants of DySuse framework. 
For all variants of our proposed DySuse framework, {we vary the layer number of self-attention in [1, 3, 5, 7]}. 
For CoupledGNN, a 3-layer structure is adopted {as reported in \citep{b4}}, and the pre-embedded dimension {is adjusted among [8, 16, 32]}. For GCN and GraphSAGE, we set the layer number and the number of hidden units to 3 and 64, {respectively, same as CoupledGNN for a fair comparison}. For MONSTOR, we stack three 2-layer GCNs and set the dimension of the hidden layer to 128 {which is the same as the parameter settings used in \citep{b55}}. For the propagation scheme, {we vary the iterative value of $q$ in [5, 7, 9]. We find that when the value of $q$ is below 5, the performance of DySuseP is extremely poor, and when the value of $q$ is 5, the effect is best in most scenarios with various inputs. We will discuss the effects of the important parameters in Section \ref{parameter_analysis} using DySuseC as the representative.}

{\textbf{Training.}
The reasoning type of DySuse depends on its Structural Feature Module, so we adopt different training methods for different variants of DySuse. 
For the transductive methods, including DySuseG and other dynamic graph embedding methods (i.e., EvolveGCN, DySAT, and TSGNet), we train on five datasets. 
For the inductive methods, including DySuseS, DySuseM, DySuseP, and DySuseC, we create a synthetic dataset BA \citep{b46} with 100 nodes. The seed size is chosen from [5, 10, 15, 20, 25], and we randomly select 20 seed sets for each seed size to generate the ground truth. Then, we train models on BA and test them on the other five datasets.}

\subsection{Experimental Results}
In this section, we test the performance of the proposed framework and compare it with several baselines from multiple dimensions. Moreover, we analyze the training time and test time. Finally, we conduct the ablation study and parameter analysis.

\begin{table*}[t]
\caption{The MAE value of different methods for susceptibility estimation on Enron dataset in the LT model.}
\label{LT}
\resizebox{2\columnwidth}{!}{
\begin{tabular}{p{1.3cm}| p{1.4cm}<{\centering}| p{1.5cm}<{\centering} p{1.4cm}<{\centering} p{1.4cm}<{\centering} p{1.4cm}<{\centering} p{1.4cm}<{\centering} p{1.4cm}<{\centering} p{1.4cm}<{\centering} p{1.4cm}<{\centering} }
\hline
\multicolumn{1}{l}{Dataset} & \begin{tabular}[c]{@{}c@{}}Seed Size\end{tabular} & EvolveGCN & DySAT & TSGNet &
\begin{tabular}[c]{@{}c@{}}DySuseG\end{tabular} & \begin{tabular}[c]{@{}c@{}}DySuseS\end{tabular} & \begin{tabular}[c]{@{}c@{}}DySuseM\end{tabular} & \begin{tabular}[c]{@{}c@{}}DySuseP\end{tabular} & \begin{tabular}[c]{@{}c@{}}DySuseC\end{tabular} \\
\hline
\hline
\multirow{5}{*}{Enron} 
& 50 & 0.484 & 0.377 & 0.454 & 0.334& 0.309& 0.200 & 0.132 & \textbf{0.067} \\
&100 & 0.467 & 0.376 & 0.422 & 0.251 & 0.338 & 0.234 & 0.117 & \textbf{0.063} \\
&150 & 0.584 & 0.366 & 0.415 & 0.323 & 0.296 & 0.265 & 0.103 & \textbf{0.051} \\
&200 & 0.617 & 0.361 & 0.400 & 0.339 & 0.285 & 0.233 & 0.141 & \textbf{0.054} \\
&250 & 0.421 & 0.353 & 0.395 & 0.293 & 0.269 & 0.227 & 0.107 & \textbf{0.050} \\
\hline
\end{tabular}}
\end{table*}

\begin{table*}[!h]
\caption{The MAE value of different methods for susceptibility estimation on Enron dataset in the TR model.}
\label{TR}
\resizebox{2\columnwidth}{!}{
\begin{tabular}{p{1.3cm}| p{1.4cm}<{\centering}| p{1.5cm}<{\centering} p{1.4cm}<{\centering} p{1.4cm}<{\centering} p{1.4cm}<{\centering} p{1.4cm}<{\centering} p{1.4cm}<{\centering} p{1.4cm}<{\centering} p{1.4cm}<{\centering} }
\hline
\multicolumn{1}{l}{Dataset} & \begin{tabular}[c]{@{}c@{}}Seed Size\end{tabular} & EvolveGCN & DySAT & TSGNet &
\begin{tabular}[c]{@{}c@{}}DySuseG\end{tabular} & \begin{tabular}[c]{@{}c@{}}DySuseS\end{tabular} & \begin{tabular}[c]{@{}c@{}}DySuseM\end{tabular} & \begin{tabular}[c]{@{}c@{}}DySuseP\end{tabular} & \begin{tabular}[c]{@{}c@{}}DySuseC\end{tabular} \\
\hline
\hline

\multirow{5}{*}{Enron} 
&50 & 0.582 & 0.450 & 0.529 & 0.267 & 0.319 & 0.269 & / & \textbf{0.046} \\
&100 & 0.557 & 0.439 & 0.525 & 0.283 & 0.243 & 0.272 & / & \textbf{0.036} \\
&150 & 0.537 & 0.420 & 0.521 & 0.264 & 0.338 & 0.269 & / & \textbf{0.026} \\
&200 & 0.599 & 0.413 & 0.517 & 0.338 & 0.324 & 0.294 & / & \textbf{0.024} \\
&250 & 0.566 & 0.403 & 0.513 & 0.297 & 0.281 & 0.274 & / & \textbf{0.017} \\
\hline
\end{tabular}}
\end{table*}

\subsubsection{Performance Comparison in the IC Model}

We compare all models on the five datasets in the IC model and report the comparison results in Table~\ref{baseline}. The results show that dynamic graph embedding models (i.e., EvolveGCN, DySAT, and TSGNet) perform unsatisfactorily for the susceptibility estimation task compared to our DySuse framework. We analyze the results from two categories as follows:

{\emph{1)} Dynamic graph embedding baselines.}
The disappointing results of dynamic graph embedding baselines means that current dynamic graph embedding models are unsuitable for directly solving dynamic susceptibility estimation problems. 
Among these dynamic graph embedding models, DySAT and TSGNet are comparable, and EvolveGCN performs worst. 
We attribute the inferior performance of EvolveGCN to its indirect embedding of temporal information: though it is innovative to use RNN to train the weights of GCN at different timestamps, this method cannot directly capture the dynamic information of influence spread over time.
DySAT and TSGNet have similar model architecture, i.e., they first obtain node embeddings with structural information through GNN and then refine the embeddings through RNN or self-attention to obtain temporal information. Note that the structural information embedding and the temporal information embedding in both models are independent, while our progressive mechanism couples structural information and temporal information together, enabling our DySuse framework, including all its variants, to outperform the above dynamic models in most cases. 

{\emph{2)} Variants of DySuse framework.}
For the variants of our framework, the performance of DySuseG and DySuseS is relatively not as good as other variants. Both DySuseG and DySuseS utilize basic GNNs (i.e., GCN and GraphSAGE) to implement the structural feature module in the DySuse framework.
In comparison, MONSTOR and DeepIS are two embedding models designed for static susceptibility estimation. The corresponding variants DySuseP and DySuseM achieve relatively high performance, indicating that our framework can achieve satisfactory prediction performance by integrating existing static susceptibility estimation methods. In particular, DySuseC outperforms all other variants in most cases, especially on the dataset Facebook (the MAE value fluctuates around 0.02). 
The reason behind its outstanding performance is twofold: First, the design of CoupledGNN makes full use of two critical components in the influence diffusion process: the iterative interaction between the node's susceptibility and the influence spread. Second, the progressive mechanism provides an effective way to tightly couple structural and temporal information.

\subsubsection{Performance Comparison in the LT and TR Models}
To validate that DySuse can work in other diffusion models with a satisfactory effect, we conduct experiments on the Enron dataset in the LT and TR \citep{b11} models.
It is worth noting that the propagation scheme in DeepIS does not present a feasible design for the TR model. Therefore, we omit the experiment of DySuseP in the TR model.

As Table~\ref{LT} and Table~\ref{TR} show, the DySuse framework remains to obtain competitive performance compared with other baselines in the LT and TR. 
The prediction results conform to the analysis in the IC model. Specifically, DySuseC still yields the best MAE value among other models. The MAE value of DySuseC fluctuates below 0.07 in the LT model and 0.05 in the TR model. 

\begin{figure*}[htbp]
    \centering
    \subfigure{
    \includegraphics[width=9cm]{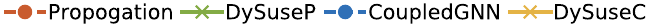}
    }
    
    \subfigure[Propagation vs DySuseP in BA.]{
    \includegraphics[width=4cm]{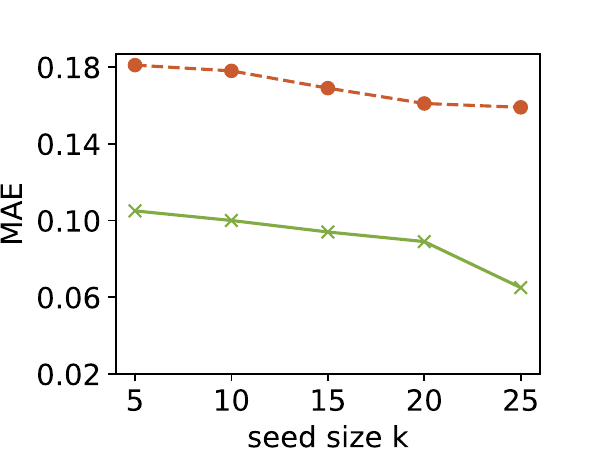}
    }
    \subfigure[CoupledGNN vs DySuseC in BA.]{
    \includegraphics[width=4cm]{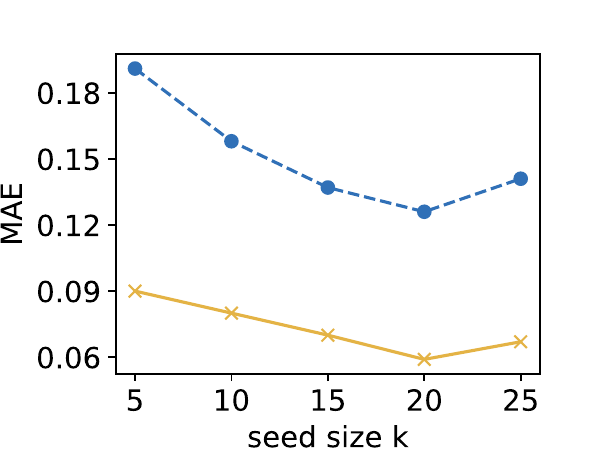}
    }
    \subfigure[Propagation vs DySuseP in Epinions.]{
    \includegraphics[width=4cm]{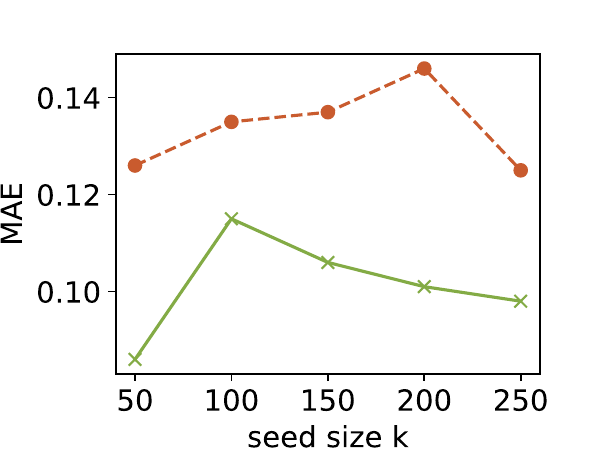}
    }
    \subfigure[CoupledGNN vs DySuseC in Epinions.]{
    \includegraphics[width=4cm]{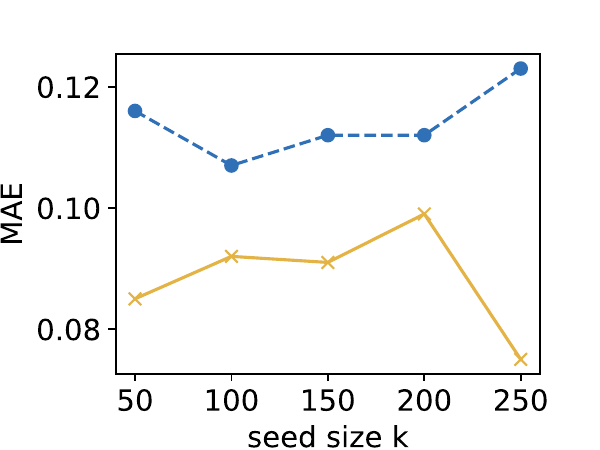}
    }
    
    \subfigure[Propagation vs DySuseP in Enron.]{
    \includegraphics[width=4cm]{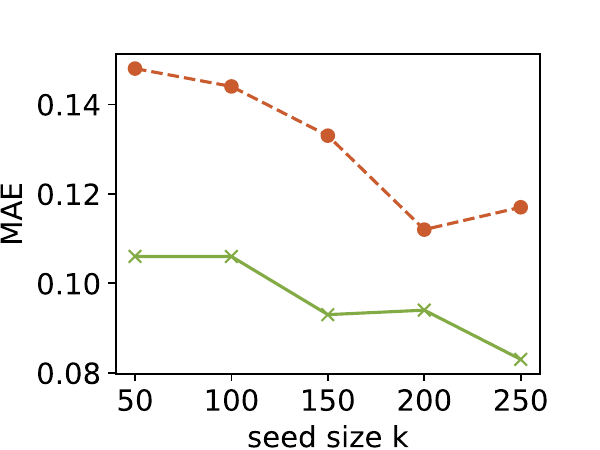}
    }
    \subfigure[CoupledGNN vs DySuseC in Enron.]{
    \includegraphics[width=4cm]{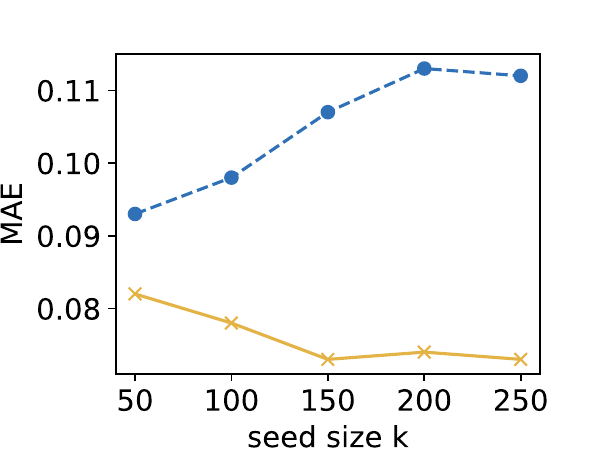}
    }
    \subfigure[Propagation vs DySuseP in Facebook.]{
    \includegraphics[width=4cm]{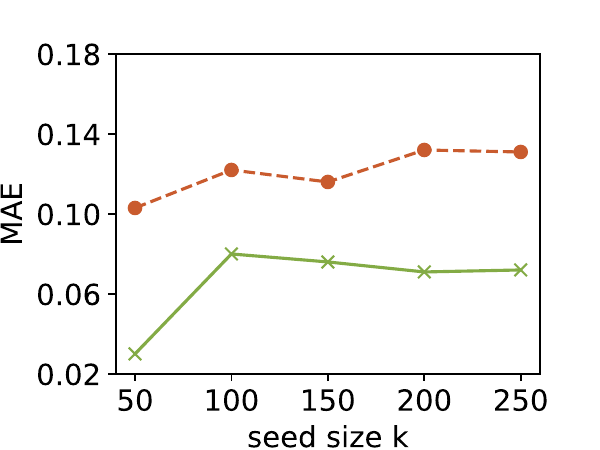}
    }
    \subfigure[CoupledGNN vs DySuseC in Facebook.]{
    \includegraphics[width=4cm]{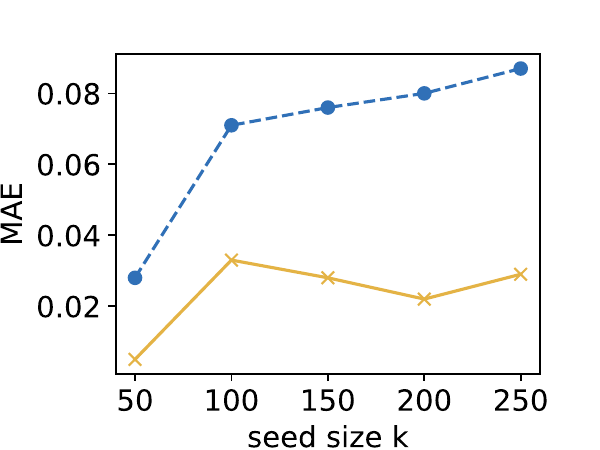}
    }
    
    \subfigure[Propagation vs DySuseP in Digg.]{
    \includegraphics[width=4cm]{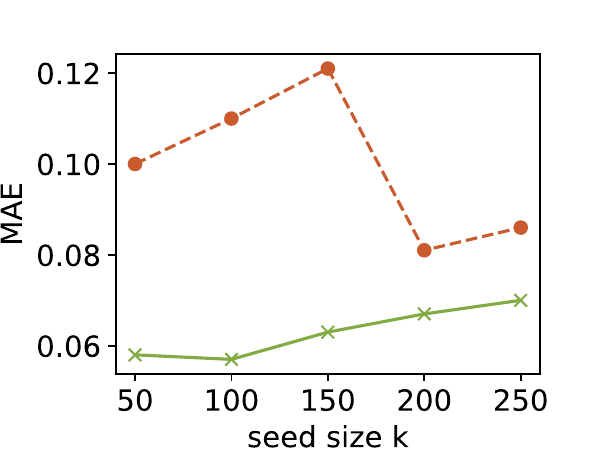}
    }
    \subfigure[CoupledGNN vs DySuseC in Digg.]{
    \includegraphics[width=4cm]{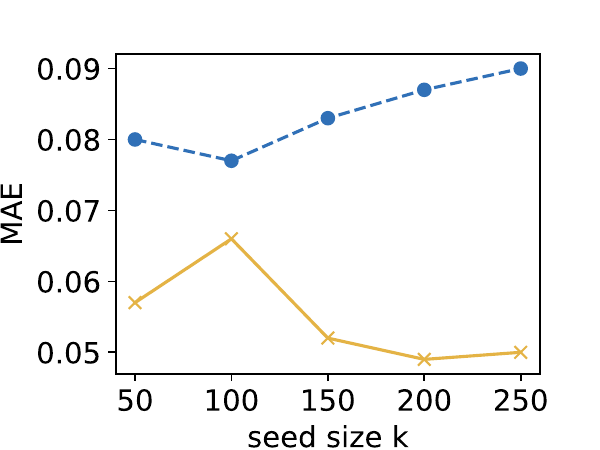}
    }
    \subfigure[Improvement of DySuseP.]{
    \includegraphics[width=4cm]{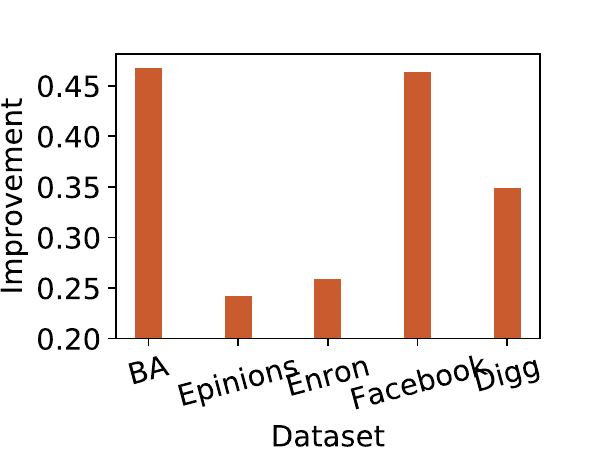}
    }
    \subfigure[Improvement of DySuseC.]{
    \includegraphics[width=4cm]{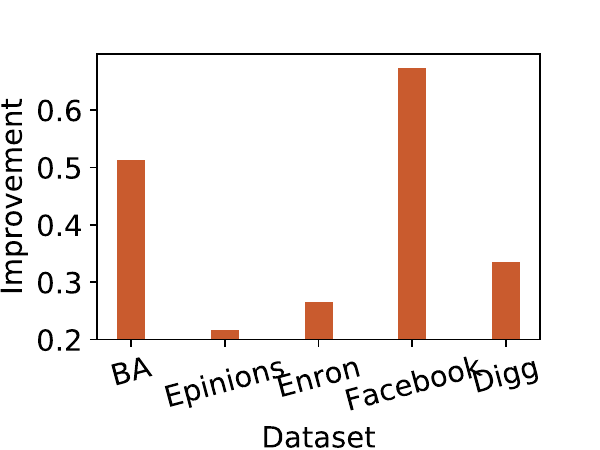}
    }
    
    \caption{{Comparison with static methods.}}
    \label{static}
\end{figure*}

\subsubsection{Comparison with Static Susceptibility Estimation Methods}
We compare our DySuse framework to some static susceptibility estimation methods (abbreviated to static methods) to analyze the benefits of utilizing temporal information in the susceptibility estimation task. The results reported in the above experiments show that DySuseP and DySuseC have relatively excellent effects. Therefore, we choose DySuseP and DySuseC as the representatives of DySuse and compare them with DeepIS and CoupledGNN, respectively. Considering that the dynamic graphs involve the deletion of nodes and edges, the static susceptibility estimation models are trained and tested on the last graph snapshot. Here we take the propagation scheme (the critical component of DeepIS) instead of DeepIS for comparison because DeepIS cannot converge when training on the last graph snapshot.

Figure~\ref{static}(a) - Figure~\ref{static}(j) show the exact MAE values of DeepIS versus DySuseP and CoupledGNN versus DySuseC on different datasets. In most cases, we can observe similar trends in the prediction performance of static methods and our framework across different seed sizes on the same dataset. Undoubtedly, the MAE values of DySuseP and DySuseC are lower than Propagation and CoupledGNN under all settings, i.e., our framework surpasses static methods for the task of susceptibility estimation. 
In order to more intuitively reveal the improvement in prediction performance of our framework compared with static methods, the average improvement of DySuseP and DySuseC are shown in Figure~\ref{static}(k) and Figure~\ref{static}(l), respectively.
The improvement of DySuseP and DySuseC on datasets BA, Facebook, and Digg is relatively significant. Numerically, DySuseP improves the propagation scheme by $46.8\%$, $46.4\%$, and $34.9\%$, while DySuseC improves CoupledGNN by $51.3\%$, $67.4\%$, and $33.5\%$ on these three datasets. We interpret that the nodes and edges in BA dataset evolve more drastically on each graph snapshot, while datasets Facebook and Digg have more timestamps. This means that the more temporal information in the graph, the more pronounced the advantages of DySuse over the static susceptibility estimation models. The results demonstrate that the way of dealing with temporal information adopted in DySuse substantially contributes to the good performance of susceptibility estimation.

\begin{figure}[tbp]
    \centering
    \subfigure[Testing time.]{
    \includegraphics[width=4cm]{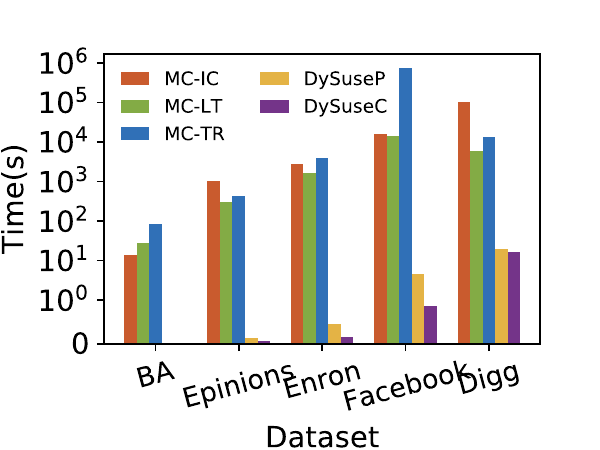}
    }
    \hspace{-0.4cm}
    \subfigure[Training time.]{
    \includegraphics[width=4cm]{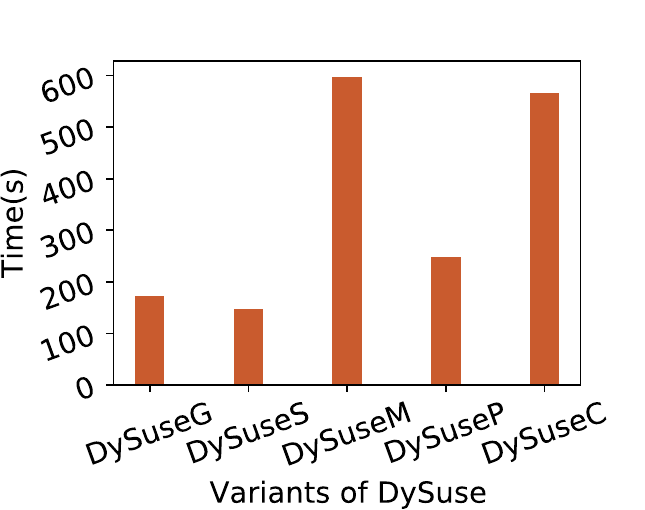}
    }
    \caption{Testing and training time.}
    \label{time}
\end{figure}
\subsubsection{Testing Time Comparison}
We present the comparisons against the calculation time of MC simulation about the testing time of DySuse in three diffusion models, i.e., IC, LT, and TR. Considering that the testing time has no correlation with the diffusion models and varies slightly among different variants, here we take the results of DySuseP and DySuseC in the IC model as the representatives of DySuse for comparison with MC simulation. For MC simulation, we repeat the simulation process 1000 times in each diffusion model, and the average results are shown in Figure~\ref{time}(a). DySuse has an overwhelming advantage in time. Specifically, DySuse is at least two orders of magnitude faster than MC simulation in each diffusion model. DySuse is even $4\sim6$ orders of magnitude faster than MC simulation on datasets BA, Enron, and Facebook. Note that the results of DySuseP and DySuseC on the dataset BA are too small (0.007s and 0.002s, respectively) to be observed even on logarithmic scale coordinates. 
{We noticed an outlier like MC-TR have high computation time on the dataset Facebook than Digg. This is because the design of the TR model is strongly related to the number of edges.}
{In addition, the different results between DySuseP and DySuseC in the same dataset in Figure~ \ref{time}(a) are mainly due to the difference in the Structural Feature Module.}
{Besides, different inputs for DySuse have different computational complexity. Under the premise that the model structure and hyperparameters are consistent, the model is closely related to the size of the dataset (including the number of nodes and edges). 
Therefore, the testing time of both DySuseP and DySuseC increases as the number of nodes or edges in the dataset increases. }

\subsubsection{Training Time Comparison}
We compare the training time of different variants of DySuse in the IC model on the dataset BA. Figure~\ref{time}(b) shows the training time results. The time consumption is determined by two primary factors: the model complexity and the astringency. 
DySuseG and DySuseS are quicker than other variants since their structural feature modules adopt a 3-layer structure. Moreover, the GCN layer and GraphSAGE layer have relatively basic architectures. DySuseM is slower than other variants because the structural feature module of DySuseM is essentially a stack of multiple GCNs, and each GCN has multiple layers. The stacking structure significantly increases DySuseM's computational complexity. The design of DySuseP's structural feature module is relatively simple. However, since the iterative value $q$ in its propagation scheme is fixed, the convergence speed is relatively slow. The structural feature module of DySuseC is implemented by two coupled GNNs. The interaction between two GNNs increases the complexity of model training. Though DySuseC is not the fastest method, it consistently performs best in estimating susceptibility in most cases.


\begin{figure}[tbp]
    \centering
    \subfigure[MAE.]{
    \includegraphics[width=3.9cm]{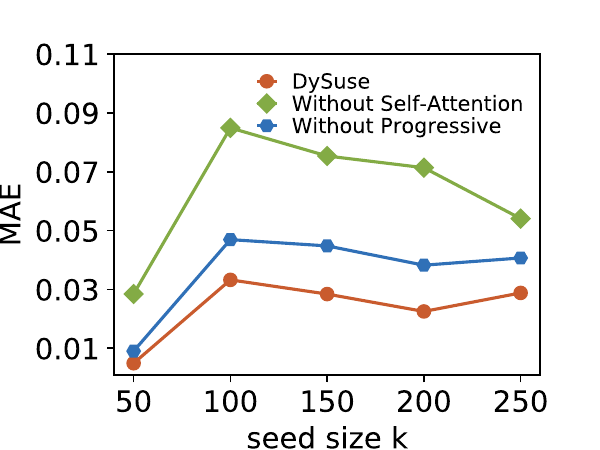}
    }
    \hspace{-0.4cm}
    \subfigure[DySuseC.]{
    \includegraphics[width=3.9cm]{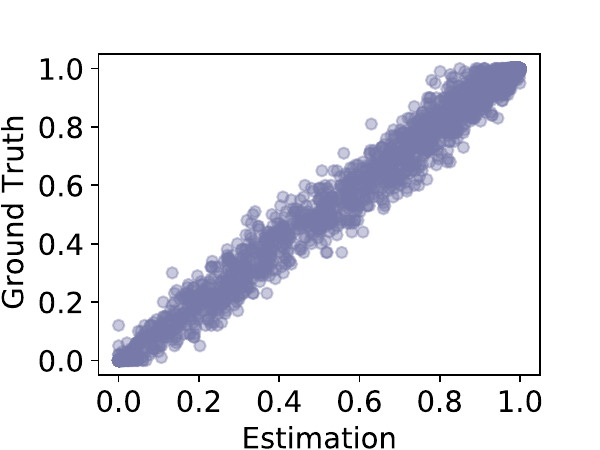}
    }
    \subfigure[DySuseC without self-attention.]{
    \includegraphics[width=3.9cm]{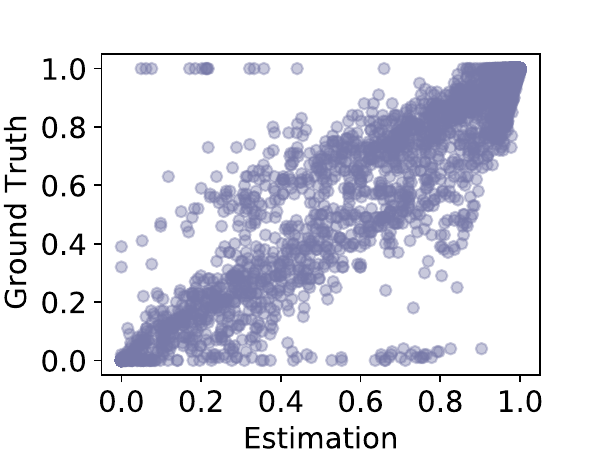}
    }
    \hspace{-0.4cm}
    \subfigure[DySuseC without progressive.]{
    \includegraphics[width=3.9cm]{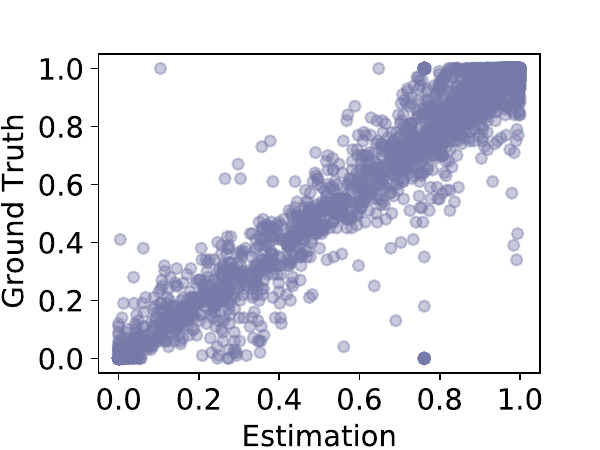}
    }
    \caption{{Ablation study for verifying the effectiveness of the self-attention block and the progressive mechanism in DySuseC.}}
    \label{ablation}
\end{figure}

\subsubsection{Ablation Study}
We conduct an ablation study to verify the effectiveness of two components in the DySuse framework, i.e., the self-attention block and the progressive mechanism. Here we take variant DySuseC as the example to illustrate its ablation study on the Facebook dataset.

In Figure~\ref{ablation}(a), we demonstrate the MAE values of DySuse and DySuse without either the progressive mechanism or the self-attention block. The results show that incorporating the progressive mechanism and the self-attention block significantly improves the prediction performance of DySuse. The self-attention block can flexibly weigh historical timestamps to fuse the susceptibility feature produced by historical information. Removing the block leads to a significant degradation in performance, i.e., $64.2\%$ on average.
Besides, the progressive mechanism also improves the performance by at least $36.3\%$.

To illustrate the comparison intuitively, we visualize an example with the settings of the dataset Facebook with 200 seeds and present Figure~\ref{ablation}(b) - Figure~\ref{ablation}(d) to show its relation between the estimated susceptibility and the ground truth. 
By observing Figure~\ref{ablation}(c) and Figure~\ref{ablation}(d), we find that both the progressive mechanism and the self-attention block can enable the DySuse framework to converge to the diagonal. Specifically, for the progressive mechanism, its performance in the middle part is not particularly ideal, while for the self-attention block, its performance in the middle part approximates the optimal results, which is largely due to the block's ability to capture valid information from the time dimension.
Overall, the mutual cooperation between the self-attention block and the progressive mechanism can obtain the most significant benefit and enable our framework to achieve excellent performance, which is verified in Figure~\ref{ablation}(b).

\begin{figure}[tbp]
    \centering
    \subfigure[Pre-emb method.]{
    \includegraphics[width=4cm]{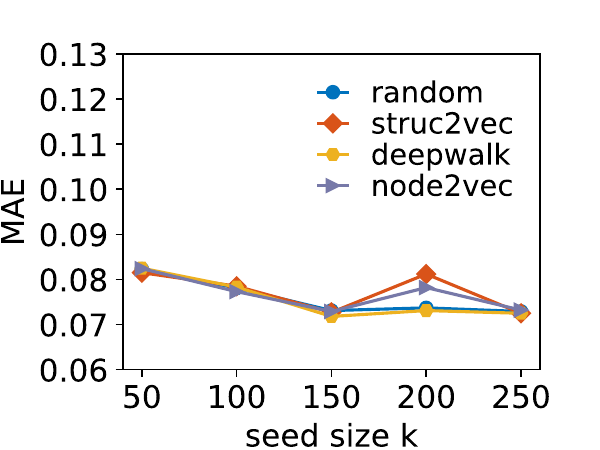}
    }
    \hspace{-3mm}
    \subfigure[{Training time.}]{
    \includegraphics[width=4cm]{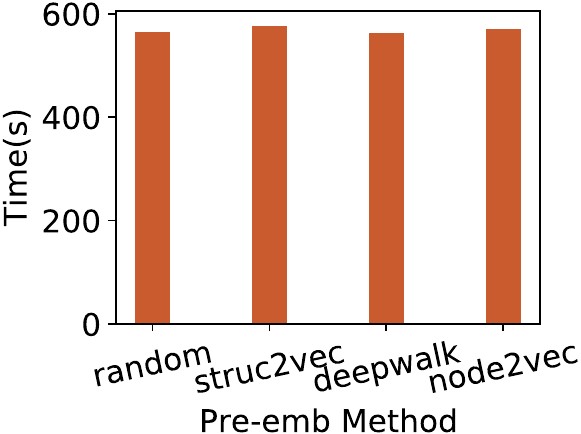}
    }
    \subfigure[Pre-emb dimension.]{
    \includegraphics[width=4cm]{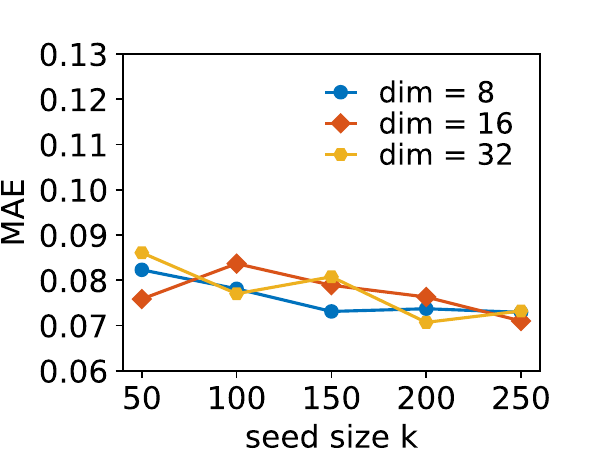}
    }
    \hspace{-3mm}
    \subfigure[{Training time.}]{
    \includegraphics[width=4cm]{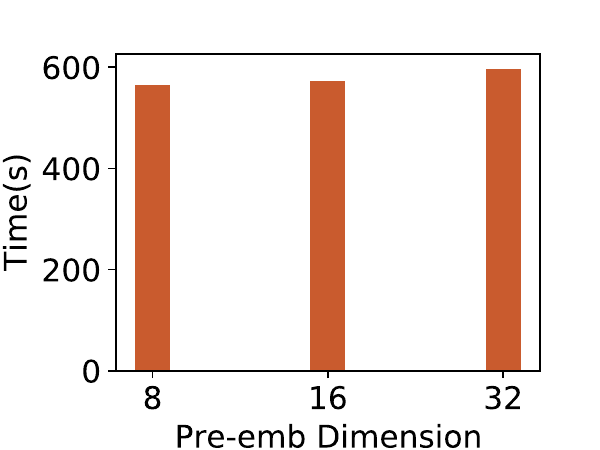}
    }
    \subfigure[Layer number.]{
    \includegraphics[width=4cm]{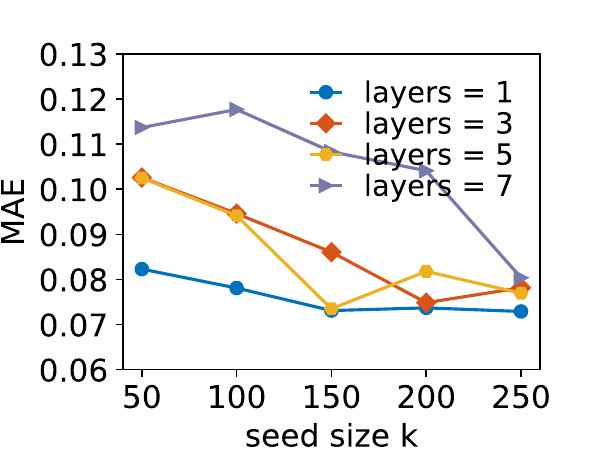}
    }
    \hspace{-3mm}
    \subfigure[{Training time.}]{
    \includegraphics[width=4cm]{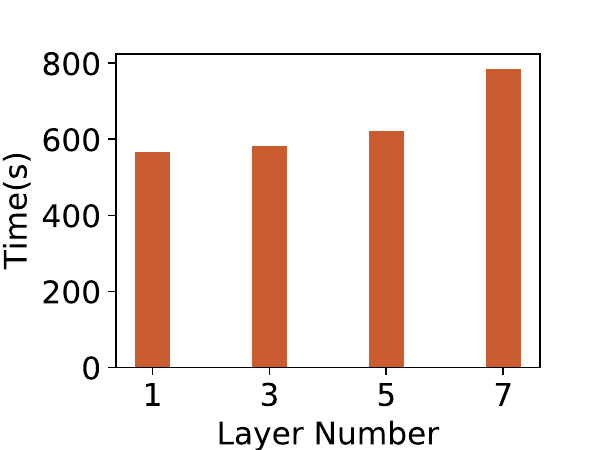}
    }
    \caption{Parameter analysis in DySuseC}
    \label{parameter}
\end{figure}

\subsubsection{Parameter Analysis}
\label{parameter_analysis}
We conduct the parameter analysis to investigate how the prediction performance {and training time} varies with some parameters in DySuse.
{We choose Enron dataset as the representative to report the results, and other datasets have similar effects. 
Here we still take DySuseC for illustration.}
For two coupled GNNs of CoupledGNN, one is used to capture node susceptibility, and the other is designed to capture influence spread. We construct the initial feature vectors for the initial node susceptibility as described in Section~\ref{input}. For the initial influence spread, we construct pre-embedding vectors and discuss the pre-embedding methods and dimensions of the pre-embedded vectors in this section.
As Figure~\ref{parameter}(a) and Figure~\ref{parameter}(c) show, the performance of DySuseC is not sensitive to the pre-embedding methods and obtains satisfactory prediction performance even when the pre-embedding dimension is low. 
{For the computation efficiency, as shown in Figure~\ref{parameter}(b) and Figure~\ref{parameter}(d), there is a slight increase in training time as the pre-embedding dimension increases. Additionally, the method of pre-embedding has a weak impact on training time.}
To optimize computation efficiency {and ease preprocessing work}, we choose the random way as the pre-embedding method and set the embedding dimension to 8 by default. 
As Figure~\ref{parameter}(c) shows, in general scenarios, the model's prediction performance gradually deteriorates as the number of self-attention layers increases. Due to the over-smooth problem of self-attention, when the number of self-attention layers increases, the information in the temporal domain cannot be captured effectively, and the quality of predicted susceptibilities is undesirable. 
{Moreover, as Figure~\ref{parameter}(f) shows, training time has a clear upward trend as the number of self-attention layers increases. Therefore, to avoid the over-smooth problem and improve training efficiency,} we set the layer number to 1 by default.

\subsubsection{Qualitative Case Study}

{In this section, we demonstrate the effectiveness of our proposed framework DySuse through a qualitative case study. We still take DySuseC as the representative of DySuse to illustrate. This case study is conducted from the end-users' point of view and primarily aims to identify nodes with high influence susceptibilities in dynamic social networks based on given influential nodes (seed nodes). We randomly select 50 nodes as influential nodes for each dataset except the smallest dataset BA, and we select 5 seed nodes for BA. We conduct susceptibility estimation by MC and DySuseC, where the results obtained from MC still act as the ground truth. The results are shown in Table \ref{qualitative}, where we present the sets of top-10 representative nodes identified by MC and DySuseC on different datasets, respectively. We find that the nodes identified by MC and DySuseC have a significant overlap nearly on each dataset. More interestingly, most of the top-ranked nodes in ground truth are also included in the set identified by DySuseC, demonstrating the effectiveness of DySuse for practical applications.}

\begin{table*}[t]
\renewcommand{\arraystretch}{1.5}
\caption{{Qualitative case study for verifying the effectiveness of DySuse framework  from the end-users' point of view, where bold identification numbers indicate the overlap between two sets obtained by MC and DySuseC, respectively.}}
\resizebox{2\columnwidth}{!}{
\begin{tabular}{p{1.3cm}| p{2.2cm}<{\centering}| p{11.3cm}<{\centering} }
\hline
\multicolumn{1}{l}{Dataset} & \begin{tabular}[c]{@{}c@{}} Method \end{tabular} & \begin{tabular}[c]{@{}c@{}} Top-10 Node No.\end{tabular} \\
\hline
\hline

\multirow{2}{*}{BA} 
& MC & \textbf{\#24} \#11 \#15 \textbf{\#23} \textbf{\#41} \#32 \#25 \textbf{\#30} \#31 \#33 \\
\cline{2-3}
& DySuseC & \#6 \#8 \#16 \textbf{\#23} \textbf{\#24} \#27 \textbf{\#30} \#37 \#40 \textbf{\#41} \\

\hline

\multirow{2}{*}{Epinions} 
& MC & \textbf{\#42} \textbf{\#199} \#445 \textbf{\#524} \textbf{\#687} \#633 \textbf{\#639} \#591 \textbf{\#688} \#779 \#807 \\
\cline{2-3}
& DySuseC & \textbf{\#42} \#152 \textbf{\#688} \#57 \textbf{\#639} \textbf{\#524} \textbf{\#687} \textbf{\#199} \#494 \#1019 \#507 \\

\hline

\multirow{2}{*}{Enron} 
& MC & \textbf{\#82} \#143 \textbf{\#260} \textbf{\#772} \#516 \textbf{\#423} \textbf{\#596} \#505 \#925 \#1062 \\
\cline{2-3}
& DySuseC & \textbf{\#423} \textbf{\#82} \textbf{\#260} \#490 \textbf{\#596} \#602 \#614 \#642 \#697 \textbf{\#772} \\

\hline

\multirow{2}{*}{Facebook}
& MC & \textbf{\#580} \textbf{\#955} \textbf{\#1071} \#1372 \#1896 \#1979 \#2142 \textbf{\#2172} \#2305 \#2464 \\
\cline{2-3}
& DySuseC & \textbf{\#580} \textbf{\#1071} \#660 \#871 \textbf{\#2172} \#1763 \textbf{\#955} \#979 \#1175 \#1261 \\

\hline

\multirow{2}{*}{Digg}
& MC & \textbf{\#6} \#38 \#25 \textbf{\#41} \textbf{\#120} \#65 \#151 \#130 \textbf{\#143} \#133 \\
\cline{2-3}
& DySuseC & \#36 \textbf{\#41} \#112 \#83 \textbf{\#143} \textbf{\#6} \#166 \#214 \#178 \textbf{\#120} \\
\hline
\end{tabular}}
\label{qualitative}
\end{table*}

\subsubsection{Quantitative Case Study}
\begin{figure}
    \centering
    \includegraphics[width=6cm]{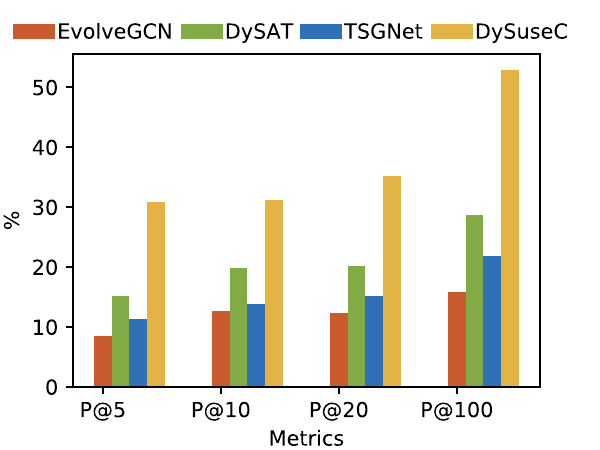}
    \caption{{Quantitative case study.}}
    \label{ex-qualitative}
\end{figure}

{In this section, we conduct a quantitative evaluation of the effectiveness of DySuse through a case study. 
In this case study, we adopt different baselines to identify nodes with top-$k$ influence susceptibilities in the Facebook dataset and still take DySuseC as the representative of DySuse. 
We use classic evaluation metrics Precision@5 (P@5), P@10, P@20, and P@100 \citep{10.1145/1008992.1009000} to evaluate the performance of different methods.
As Figure~ \ref{ex-qualitative} demonstrates, DySuseC's performance is superior to the other baselines on all metrics. Specifically, the performance of all methods gets improved with varying degrees as $k$ increases. 
Especially for DySuseC, the value of P@100 has significant progress compared to P@5, P@10, and P@20. 
This is because P@100 considers a larger number of top-ranked nodes and gives all methods more opportunities to identify relevant top-ranked nodes.
For the application of the susceptibility estimation task, a low value of $k$ often has no practical significance. Taking a real-world marketing activity as an example, we always try to target an advertisement at a large number of users who may be influenced rather than just a few top-ranked users.
Therefore, Precision@100 can be a more comprehensive measure of the effectiveness of our methods than P@5, P@10, and P@20, as it takes into account a larger number of nodes.}

\section{Conclusions and Future Work}\label{GG}
This paper investigated susceptibility estimation in dynamic networks and proposed the DySuse framework. It can predict susceptibility by tightly coupling structural and temporal information. Specifically, for the topological information of influence diffusion, we independently captured it on each single graph snapshot through a structural feature module. Meanwhile, to tightly couple the topological and temporal information, we invented a progressive mechanism to strengthen the dependence between the initial feature vectors at different timestamps. Based on that, we adopted the self-attention block to further capture temporal dependency by flexibly weighting historical timestamps. 

We performed extensive experiments on multiple widely used datasets. 
The experiments show that our framework achieved remarkable results in multiple diffusion models, including IC, LT and TR, so that our framework can adapt to multiple diffusion models with great estimation performance. 
As a result, DySuse not only outperforms existing dynamic graph embedding models in terms of prediction performance in most cases, but it is undoubtedly far better than the traditional MC simulation solution in calculation speed.
This reveals that our framework has significant practical application value. {Inspired by Zhang et al. \citep{9970372}, we envision that DySuse is a general framework for some influence analysis tasks, such as influence spread estimation and influence contribution estimation in dynamic networks. The proposal of DySuse effectively fills the research gap in node-level influence analysis in dynamic social networks.}

{Inspired by Zhang et al. \citep{ZHANG20208925}, we realize a limitation in our framework DySuse, i.e., the performance of DySuse may be influenced by the Structural Feature Module. In the future, there are several directions that we believe are worth exploring to improve the performance of the framework:
\begin{itemize}
\item We are interested in designing a module with tightly coupling with DySuse for the Structural Feature Module in DySuse. We attempt to draw inspiration from the random walk and link it to the process of influence diffusion in dynamic social networks to capture the structure information.
\item It is promising to study how to incorporate prior information of our task into DySuse using existing technologies. For example, by using event detection technology \citep{7314900,6525357,misleading_Detection} to identify specific relevant events, we can estimate users' susceptibilities more accurately for specific events. This enables the model to solve more practical problems, such as identifying high-susceptibility users in dynamic networks for rumor propagation, controlling the development of public opinion, and product marketing.
\end{itemize}}

\printcredits

\section*{Declaration of competing interest}
The authors declare that they have no known competing financial interests or personal relationships that could have appeared to influence the work reported in this paper.

\section*{Acknowledgments}
This work was partially supported by the National Natural Science Foundation of China (61972272), the Natural Science Foundation of the Jiangsu Higher Education Institutions of China (21KJA520008), Postgraduate Research \& Practice Innovation Program of Jiangsu Province (SJCX23\_1660), Qinlan Project of Jiangsu Province of China, and Project Funded by the Priority Academic Program Development of Jiangsu Higher Education Institutions.

\bibliographystyle{cas-model2-names}

\bibliography{mybibliography}

\begin{thebibliography}{43}
\expandafter\ifx\csname natexlab\endcsname\relax\def\natexlab#1{#1}\fi
\providecommand{\url}[1]{\texttt{#1}}
\providecommand{\href}[2]{#2}
\providecommand{\path}[1]{#1}
\providecommand{\DOIprefix}{doi:}
\providecommand{\ArXivprefix}{arXiv:}
\providecommand{\URLprefix}{URL: }
\providecommand{\Pubmedprefix}{pmid:}
\providecommand{\doi}[1]{\href{http://dx.doi.org/#1}{\path{#1}}}
\providecommand{\Pubmed}[1]{\href{pmid:#1}{\path{#1}}}
\providecommand{\bibinfo}[2]{#2}
\ifx\xfnm\relax \def\xfnm[#1]{\unskip,\space#1}\fi
\bibitem[{Aiello et~al.(2013)Aiello, Petkos, Martin, Corney, Papadopoulos,
  Skraba, Göker, Kompatsiaris and Jaimes}]{6525357}
\bibinfo{author}{Aiello, L.M.}, \bibinfo{author}{Petkos, G.},
  \bibinfo{author}{Martin, C.}, \bibinfo{author}{Corney, D.},
  \bibinfo{author}{Papadopoulos, S.}, \bibinfo{author}{Skraba, R.},
  \bibinfo{author}{Göker, A.}, \bibinfo{author}{Kompatsiaris, I.},
  \bibinfo{author}{Jaimes, A.}, \bibinfo{year}{2013}.
\newblock \bibinfo{title}{Sensing trending topics in twitter}.
\newblock \bibinfo{journal}{IEEE Transactions on Multimedia}
  \bibinfo{volume}{15}, \bibinfo{pages}{1268--1282}.
\bibitem[{Boididou et~al.(2018)Boididou, Papadopoulos, Zampoglou, Apostolidis,
  Papadopoulou and Kompatsiaris}]{misleading_Detection}
\bibinfo{author}{Boididou, C.}, \bibinfo{author}{Papadopoulos, S.},
  \bibinfo{author}{Zampoglou, M.}, \bibinfo{author}{Apostolidis, L.},
  \bibinfo{author}{Papadopoulou, O.}, \bibinfo{author}{Kompatsiaris, I.},
  \bibinfo{year}{2018}.
\newblock \bibinfo{title}{Detection and visualization of misleading content on
  twitter}.
\newblock \bibinfo{journal}{International Journal of Multimedia Information
  Retrieval} \bibinfo{volume}{7}.
\bibitem[{Borgs et~al.(2014)Borgs, Brautbar, Chayes and
  Lucier}]{DBLP:conf/soda/BorgsBCL14}
\bibinfo{author}{Borgs, C.}, \bibinfo{author}{Brautbar, M.},
  \bibinfo{author}{Chayes, J.}, \bibinfo{author}{Lucier, B.},
  \bibinfo{year}{2014}.
\newblock \bibinfo{title}{Maximizing social influence in nearly optimal time},
  in: \bibinfo{booktitle}{Proceedings of the Twenty-Fifth Annual ACM-SIAM
  Symposium on Discrete Algorithms}, \bibinfo{publisher}{Society for Industrial
  and Applied Mathematics}, \bibinfo{address}{USA}. pp.
  \bibinfo{pages}{946--957}.
\bibitem[{Buckley and Voorhees(2004)}]{10.1145/1008992.1009000}
\bibinfo{author}{Buckley, C.}, \bibinfo{author}{Voorhees, E.M.},
  \bibinfo{year}{2004}.
\newblock \bibinfo{title}{Retrieval evaluation with incomplete information},
  in: \bibinfo{booktitle}{Proceedings of the 27th Annual International ACM
  SIGIR Conference on Research and Development in Information Retrieval},
  \bibinfo{publisher}{Association for Computing Machinery}. pp.
  \bibinfo{pages}{25--32}.
\bibitem[{Cao et~al.(2020)Cao, Shen, Gao, Wei and Cheng}]{b4}
\bibinfo{author}{Cao, Q.}, \bibinfo{author}{Shen, H.}, \bibinfo{author}{Gao,
  J.}, \bibinfo{author}{Wei, B.}, \bibinfo{author}{Cheng, X.},
  \bibinfo{year}{2020}.
\newblock \bibinfo{title}{Popularity prediction on social platforms with
  coupled graph neural networks}, in: \bibinfo{booktitle}{Proceedings of the
  13th International Conference on Web Search and Data Mining},
  \bibinfo{publisher}{Association for Computing Machinery},
  \bibinfo{address}{New York, NY, USA}. pp. \bibinfo{pages}{70--78}.
\bibitem[{Chen and Wong(2021)}]{DBLP:conf/wsdm/ChenW21}
\bibinfo{author}{Chen, T.}, \bibinfo{author}{Wong, R.C.W.},
  \bibinfo{year}{2021}.
\newblock \bibinfo{title}{An efficient and effective framework for
  session-based social recommendation}, in: \bibinfo{booktitle}{Proceedings of
  the 14th ACM International Conference on Web Search and Data Mining},
  \bibinfo{publisher}{Association for Computing Machinery},
  \bibinfo{address}{New York, NY, USA}. pp. \bibinfo{pages}{400--408}.
\bibitem[{Chen et~al.(2009)Chen, Wang and Yang}]{b20}
\bibinfo{author}{Chen, W.}, \bibinfo{author}{Wang, Y.}, \bibinfo{author}{Yang,
  S.}, \bibinfo{year}{2009}.
\newblock \bibinfo{title}{Efficient influence maximization in social networks},
  in: \bibinfo{booktitle}{Proceedings of the 15th ACM SIGKDD International
  Conference on Knowledge Discovery and Data Mining}, pp.
  \bibinfo{pages}{199--208}.
\bibitem[{De~Choudhury et~al.(2009)De~Choudhury, Sundaram, John and
  Seligmann}]{b50}
\bibinfo{author}{De~Choudhury, M.}, \bibinfo{author}{Sundaram, H.},
  \bibinfo{author}{John, A.}, \bibinfo{author}{Seligmann, D.D.},
  \bibinfo{year}{2009}.
\newblock \bibinfo{title}{Social synchrony: Predicting mimicry of user actions
  in online social media}, in: \bibinfo{booktitle}{2009 International
  Conference on Computational Science and Engineering}, pp.
  \bibinfo{pages}{151--158}.
\bibitem[{Doulamis et~al.(2016)Doulamis, Doulamis, Kokkinos and
  Varvarigos}]{7314900}
\bibinfo{author}{Doulamis, N.D.}, \bibinfo{author}{Doulamis, A.D.},
  \bibinfo{author}{Kokkinos, P.}, \bibinfo{author}{Varvarigos, E.M.},
  \bibinfo{year}{2016}.
\newblock \bibinfo{title}{Event detection in twitter microblogging}.
\newblock \bibinfo{journal}{IEEE Transactions on Cybernetics}
  \bibinfo{volume}{46}, \bibinfo{pages}{2810--2824}.
\bibitem[{Gruhl et~al.(2004)Gruhl, Guha, Liben-Nowell and
  Tomkins}]{10.1145/988672.988739}
\bibinfo{author}{Gruhl, D.}, \bibinfo{author}{Guha, R.},
  \bibinfo{author}{Liben-Nowell, D.}, \bibinfo{author}{Tomkins, A.},
  \bibinfo{year}{2004}.
\newblock \bibinfo{title}{Information diffusion through blogspace}, in:
  \bibinfo{booktitle}{Proceedings of the 13th International Conference on World
  Wide Web}, \bibinfo{publisher}{Association for Computing Machinery},
  \bibinfo{address}{New York, NY, USA}. pp. \bibinfo{pages}{491--501}.
\bibitem[{Guo and Wu(2020)}]{10.1145/3399661}
\bibinfo{author}{Guo, J.}, \bibinfo{author}{Wu, W.}, \bibinfo{year}{2020}.
\newblock \bibinfo{title}{Influence maximization: Seeding based on community
  structure}.
\newblock \bibinfo{journal}{ACM Trans. Knowl. Discov. Data}
  \bibinfo{volume}{14}.
\bibitem[{Guo et~al.(2020)Guo, Wang, Wei and Chen}]{10.1145/3318464.3389740}
\bibinfo{author}{Guo, Q.}, \bibinfo{author}{Wang, S.}, \bibinfo{author}{Wei,
  Z.}, \bibinfo{author}{Chen, M.}, \bibinfo{year}{2020}.
\newblock \bibinfo{title}{Influence Maximization Revisited: Efficient Reverse
  Reachable Set Generation with Bound Tightened}.
\newblock pp. \bibinfo{pages}{2167--2181}.
\bibitem[{Hamilton et~al.(2017)Hamilton, Ying and Leskovec}]{b54}
\bibinfo{author}{Hamilton, W.L.}, \bibinfo{author}{Ying, R.},
  \bibinfo{author}{Leskovec, J.}, \bibinfo{year}{2017}.
\newblock \bibinfo{title}{Inductive representation learning on large graphs},
  in: \bibinfo{booktitle}{Proceedings of the 31st International Conference on
  Neural Information Processing Systems}, \bibinfo{publisher}{Curran Associates
  Inc.}, \bibinfo{address}{Red Hook, NY, USA}. pp. \bibinfo{pages}{1025--1035}.
\bibitem[{Jung et~al.(2012)Jung, Heo and Chen}]{b21}
\bibinfo{author}{Jung, K.}, \bibinfo{author}{Heo, W.}, \bibinfo{author}{Chen,
  W.}, \bibinfo{year}{2012}.
\newblock \bibinfo{title}{Irie: Scalable and robust influence maximization in
  social networks}, in: \bibinfo{booktitle}{2013 IEEE 13th International
  Conference on Data Mining}, \bibinfo{publisher}{IEEE Computer Society},
  \bibinfo{address}{Los Alamitos, CA, USA}. pp. \bibinfo{pages}{918--923}.
\bibitem[{Kempe et~al.(2003)Kempe, Kleinberg and Tardos}]{b11}
\bibinfo{author}{Kempe, D.}, \bibinfo{author}{Kleinberg, J.},
  \bibinfo{author}{Tardos, E.}, \bibinfo{year}{2003}.
\newblock \bibinfo{title}{Maximizing the spread of influence through a social
  network}, in: \bibinfo{booktitle}{Proceedings of the Ninth ACM SIGKDD
  International Conference on Knowledge Discovery and Data Mining},
  \bibinfo{publisher}{Association for Computing Machinery},
  \bibinfo{address}{New York, NY, USA}. pp. \bibinfo{pages}{137--146}.
\bibitem[{Kipf and Welling(2017)}]{b26}
\bibinfo{author}{Kipf, T.N.}, \bibinfo{author}{Welling, M.},
  \bibinfo{year}{2017}.
\newblock \bibinfo{title}{Semi-supervised classification with graph
  convolutional networks}, in: \bibinfo{booktitle}{5th International Conference
  on Learning Representations, {ICLR} 2017, Toulon, France, April 24-26, 2017,
  Conference Track Proceedings}.
\bibitem[{Klimt and Yang(2004)}]{b48}
\bibinfo{author}{Klimt, B.}, \bibinfo{author}{Yang, Y.}, \bibinfo{year}{2004}.
\newblock \bibinfo{title}{The enron corpus: A new dataset for email
  classification research}, in: \bibinfo{booktitle}{Proceedings of the 15th
  European Conference on Machine Learning},
  \bibinfo{publisher}{Springer-Verlag}, \bibinfo{address}{Berlin, Heidelberg}.
  pp. \bibinfo{pages}{217--226}.
\bibitem[{Ko et~al.(2020)Ko, Lee, Shin and Park}]{b55}
\bibinfo{author}{Ko, J.}, \bibinfo{author}{Lee, K.}, \bibinfo{author}{Shin,
  K.}, \bibinfo{author}{Park, N.}, \bibinfo{year}{2020}.
\newblock \bibinfo{title}{Monstor: An inductive approach for estimating and
  maximizing influence over unseen networks}, in: \bibinfo{booktitle}{2020
  IEEE/ACM International Conference on Advances in Social Networks Analysis and
  Mining (ASONAM)}, pp. \bibinfo{pages}{204--211}.
\bibitem[{Krizhevsky(2010)}]{b53}
\bibinfo{author}{Krizhevsky, A.}, \bibinfo{year}{2010}.
\newblock \bibinfo{title}{Convolutional deep belief networks on cifar-10}.
\bibitem[{Krizhevsky et~al.(2017)Krizhevsky, Sutskever and
  Hinton}]{NIPS2012_c399862d}
\bibinfo{author}{Krizhevsky, A.}, \bibinfo{author}{Sutskever, I.},
  \bibinfo{author}{Hinton, G.E.}, \bibinfo{year}{2017}.
\newblock \bibinfo{title}{Imagenet classification with deep convolutional
  neural networks}.
\newblock \bibinfo{journal}{Commun. ACM} \bibinfo{volume}{60},
  \bibinfo{pages}{84--90}.
\bibitem[{Massa and Avesani(2005)}]{b47}
\bibinfo{author}{Massa, P.}, \bibinfo{author}{Avesani, P.},
  \bibinfo{year}{2005}.
\newblock \bibinfo{title}{Controversial users demand local trust metrics: An
  experimental study on epinions.com community}, in:
  \bibinfo{booktitle}{Proceedings of the 20th National Conference on Artificial
  Intelligence - Volume 1}, \bibinfo{publisher}{AAAI Press}. pp.
  \bibinfo{pages}{121--126}.
\bibitem[{McAuley and Leskovec(2012)}]{b49}
\bibinfo{author}{McAuley, J.}, \bibinfo{author}{Leskovec, J.},
  \bibinfo{year}{2012}.
\newblock \bibinfo{title}{Learning to discover social circles in ego networks},
  in: \bibinfo{booktitle}{Proceedings of the 25th International Conference on
  Neural Information Processing Systems - Volume 1}, \bibinfo{publisher}{Curran
  Associates Inc.}, \bibinfo{address}{Red Hook, NY, USA}. pp.
  \bibinfo{pages}{539--547}.
\bibitem[{Mnih et~al.(2014)Mnih, Heess, Graves and
  Kavukcuoglu}]{10.5555/2969033.2969073}
\bibinfo{author}{Mnih, V.}, \bibinfo{author}{Heess, N.},
  \bibinfo{author}{Graves, A.}, \bibinfo{author}{Kavukcuoglu, K.},
  \bibinfo{year}{2014}.
\newblock \bibinfo{title}{Recurrent models of visual attention}, in:
  \bibinfo{booktitle}{Proceedings of the 27th International Conference on
  Neural Information Processing Systems - Volume 2}, \bibinfo{publisher}{MIT
  Press}, \bibinfo{address}{Cambridge, MA, USA}. pp.
  \bibinfo{pages}{2204--2212}.
\bibitem[{Nguyen et~al.(2017)Nguyen, Nguyen, Vu and Dinh}]{10.1145/3084457}
\bibinfo{author}{Nguyen, H.T.}, \bibinfo{author}{Nguyen, T.P.},
  \bibinfo{author}{Vu, T.N.}, \bibinfo{author}{Dinh, T.N.},
  \bibinfo{year}{2017}.
\newblock \bibinfo{title}{Outward influence and cascade size estimation in
  billion-scale networks}, in: \bibinfo{booktitle}{Proceedings of the 2017 ACM
  SIGMETRICS / International Conference on Measurement and Modeling of Computer
  Systems}, p.~\bibinfo{pages}{63}.
\bibitem[{Onnela et~al.(2007)Onnela, Saram\"{a}ki, Hyv\"{o}nen, Szab\'{o},
  Lazer, Kaski, Kert\'{e}sz and Barab\'{a}si}]{b46}
\bibinfo{author}{Onnela, J.P.}, \bibinfo{author}{Saram\"{a}ki, J.},
  \bibinfo{author}{Hyv\"{o}nen, J.}, \bibinfo{author}{Szab\'{o}, G.},
  \bibinfo{author}{Lazer, D.}, \bibinfo{author}{Kaski, K.},
  \bibinfo{author}{Kert\'{e}sz, J.}, \bibinfo{author}{Barab\'{a}si, A.L.},
  \bibinfo{year}{2007}.
\newblock \bibinfo{title}{Structure and tie strengths in mobile communication
  networks}.
\newblock \bibinfo{journal}{Proceedings of the National Academy of Sciences}
  \bibinfo{volume}{104}, \bibinfo{pages}{7332--7336}.
\bibitem[{Panagopoulos et~al.(2022)Panagopoulos, Malliaros and
  Vazirgiannis}]{b7}
\bibinfo{author}{Panagopoulos, G.}, \bibinfo{author}{Malliaros, F.D.},
  \bibinfo{author}{Vazirgiannis, M.}, \bibinfo{year}{2022}.
\newblock \bibinfo{title}{Multi-task learning for influence estimation and
  maximization}.
\newblock \bibinfo{journal}{IEEE Transactions on Knowledge and Data
  Engineering} \bibinfo{volume}{34}, \bibinfo{pages}{4398--4409}.
\bibitem[{Pareja et~al.(2020)Pareja, Domeniconi, Chen, Ma, Suzumura, Kanezashi,
  Kaler, Schardl and Leiserson}]{b34}
\bibinfo{author}{Pareja, A.}, \bibinfo{author}{Domeniconi, G.},
  \bibinfo{author}{Chen, J.}, \bibinfo{author}{Ma, T.},
  \bibinfo{author}{Suzumura, T.}, \bibinfo{author}{Kanezashi, H.},
  \bibinfo{author}{Kaler, T.}, \bibinfo{author}{Schardl, T.B.},
  \bibinfo{author}{Leiserson, C.E.}, \bibinfo{year}{2020}.
\newblock \bibinfo{title}{Evolvegcn: Evolving graph convolutional networks for
  dynamic graphs}, in: \bibinfo{booktitle}{The Thirty-Fourth {AAAI} Conference
  on Artificial Intelligence}, \bibinfo{publisher}{{AAAI} Press}. pp.
  \bibinfo{pages}{5363--5370}.
\bibitem[{Park and Neville(2019)}]{b31}
\bibinfo{author}{Park, H.}, \bibinfo{author}{Neville, J.},
  \bibinfo{year}{2019}.
\newblock \bibinfo{title}{Exploiting interaction links for node classification
  with deep graph neural networks}, in: \bibinfo{booktitle}{Proceedings of the
  28th International Joint Conference on Artificial Intelligence},
  \bibinfo{publisher}{AAAI Press}. pp. \bibinfo{pages}{3223--3230}.
\bibitem[{Perozzi et~al.(2014)Perozzi, Al-Rfou and
  Skiena}]{10.1145/2623330.2623732}
\bibinfo{author}{Perozzi, B.}, \bibinfo{author}{Al-Rfou, R.},
  \bibinfo{author}{Skiena, S.}, \bibinfo{year}{2014}.
\newblock \bibinfo{title}{Deepwalk: Online learning of social representations},
  in: \bibinfo{booktitle}{Proceedings of the 20th ACM SIGKDD International
  Conference on Knowledge Discovery and Data Mining},
  \bibinfo{publisher}{Association for Computing Machinery}. pp.
  \bibinfo{pages}{701--710}.
\bibitem[{Sankar et~al.(2020)Sankar, Wu, Gou, Zhang and Yang}]{b32}
\bibinfo{author}{Sankar, A.}, \bibinfo{author}{Wu, Y.}, \bibinfo{author}{Gou,
  L.}, \bibinfo{author}{Zhang, W.}, \bibinfo{author}{Yang, H.},
  \bibinfo{year}{2020}.
\newblock \bibinfo{title}{Dysat: Deep neural representation learning on dynamic
  graphs via self-attention networks}, in: \bibinfo{booktitle}{Proceedings of
  the 13th International Conference on Web Search and Data Mining},
  \bibinfo{publisher}{Association for Computing Machinery},
  \bibinfo{address}{New York, NY, USA}. pp. \bibinfo{pages}{519--527}.
\bibitem[{Sun et~al.(2020)Sun, Chen, Yu and Chen}]{10.1145/3336191.3371791}
\bibinfo{author}{Sun, L.}, \bibinfo{author}{Chen, A.}, \bibinfo{author}{Yu,
  P.S.}, \bibinfo{author}{Chen, W.}, \bibinfo{year}{2020}.
\newblock \bibinfo{title}{Influence maximization with spontaneous user
  adoption}, in: \bibinfo{booktitle}{Proceedings of the 13th International
  Conference on Web Search and Data Mining}, \bibinfo{publisher}{Association
  for Computing Machinery}. pp. \bibinfo{pages}{573--581}.
\bibitem[{Tang et~al.(2014)Tang, Xiao and Shi}]{b13}
\bibinfo{author}{Tang, Y.}, \bibinfo{author}{Xiao, X.}, \bibinfo{author}{Shi,
  Y.}, \bibinfo{year}{2014}.
\newblock \bibinfo{title}{Influence maximization: Near-optimal time complexity
  meets practical efficiency}, in: \bibinfo{booktitle}{Proceedings of the 2014
  ACM SIGMOD International Conference on Management of Data},
  \bibinfo{publisher}{Association for Computing Machinery},
  \bibinfo{address}{New York, NY, USA}. pp. \bibinfo{pages}{75--86}.
\bibitem[{Vaswani et~al.(2017)Vaswani, Shazeer, Parmar, Uszkoreit, Jones,
  Gomez, Kaiser and Polosukhin}]{b51}
\bibinfo{author}{Vaswani, A.}, \bibinfo{author}{Shazeer, N.},
  \bibinfo{author}{Parmar, N.}, \bibinfo{author}{Uszkoreit, J.},
  \bibinfo{author}{Jones, L.}, \bibinfo{author}{Gomez, A.N.},
  \bibinfo{author}{Kaiser, L.}, \bibinfo{author}{Polosukhin, I.},
  \bibinfo{year}{2017}.
\newblock \bibinfo{title}{Attention is all you need}, in:
  \bibinfo{booktitle}{Proceedings of the 31st International Conference on
  Neural Information Processing Systems}, \bibinfo{publisher}{Curran Associates
  Inc.}, \bibinfo{address}{Red Hook, NY, USA}. pp. \bibinfo{pages}{6000--6010}.
\bibitem[{Wu and Wang(2020)}]{DBLP:journals/tst/wu20}
\bibinfo{author}{Wu, J.}, \bibinfo{author}{Wang, N.}, \bibinfo{year}{2020}.
\newblock \bibinfo{title}{Approximating special social influence maximization
  problems}.
\newblock \bibinfo{journal}{Tsinghua Science and Technology}
  \bibinfo{volume}{25}, \bibinfo{pages}{703--711}.
\bibitem[{Wu et~al.(2022)Wu, Zhou, Liu, Li and Gu}]{9835455}
\bibinfo{author}{Wu, Z.}, \bibinfo{author}{Zhou, J.}, \bibinfo{author}{Liu,
  L.}, \bibinfo{author}{Li, C.}, \bibinfo{author}{Gu, F.},
  \bibinfo{year}{2022}.
\newblock \bibinfo{title}{Deep popularity prediction in multi-source cascade
  with heri-gcn}, in: \bibinfo{booktitle}{2022 IEEE 38th International
  Conference on Data Engineering (ICDE)}, pp. \bibinfo{pages}{1714--1726}.
\bibitem[{Xia et~al.(2021)Xia, Li, Wu and Li}]{b17}
\bibinfo{author}{Xia, W.}, \bibinfo{author}{Li, Y.}, \bibinfo{author}{Wu, J.},
  \bibinfo{author}{Li, S.}, \bibinfo{year}{2021}.
\newblock \bibinfo{title}{Deepis: Susceptibility estimation on social
  networks}, in: \bibinfo{booktitle}{Proceedings of the 14th ACM International
  Conference on Web Search and Data Mining}, \bibinfo{publisher}{Association
  for Computing Machinery}, \bibinfo{address}{New York, NY, USA}. pp.
  \bibinfo{pages}{761--769}.
\bibitem[{Yang et~al.(2019)Yang, Tang, Sun, Cui and Liu}]{b3}
\bibinfo{author}{Yang, C.}, \bibinfo{author}{Tang, J.}, \bibinfo{author}{Sun,
  M.}, \bibinfo{author}{Cui, G.}, \bibinfo{author}{Liu, Z.},
  \bibinfo{year}{2019}.
\newblock \bibinfo{title}{Multi-scale information diffusion prediction with
  reinforced recurrent networks}, in: \bibinfo{booktitle}{Proceedings of the
  28th International Joint Conference on Artificial Intelligence},
  \bibinfo{publisher}{AAAI Press}. pp. \bibinfo{pages}{4033--4039}.
\bibitem[{Yang and Leskovec(2010)}]{ICDM2010Yang}
\bibinfo{author}{Yang, J.}, \bibinfo{author}{Leskovec, J.},
  \bibinfo{year}{2010}.
\newblock \bibinfo{title}{Modeling information diffusion in implicit networks},
  in: \bibinfo{booktitle}{2010 IEEE International Conference on Data Mining},
  pp. \bibinfo{pages}{599--608}.
\bibitem[{Zhang et~al.(2020)Zhang, Li, Xiao and Zhang}]{ZHANG20208925}
\bibinfo{author}{Zhang, J.}, \bibinfo{author}{Li, Y.}, \bibinfo{author}{Xiao,
  W.}, \bibinfo{author}{Zhang, Z.}, \bibinfo{year}{2020}.
\newblock \bibinfo{title}{Non-iterative and fast deep learning: Multilayer
  extreme learning machines}.
\newblock \bibinfo{journal}{Journal of the Franklin Institute}
  \bibinfo{volume}{357}, \bibinfo{pages}{8925--8955}.
\bibitem[{Zhang et~al.(2023)Zhang, Zhao, Shone, Li, Frangi, Xie and
  Zhang}]{9970372}
\bibinfo{author}{Zhang, J.}, \bibinfo{author}{Zhao, Y.},
  \bibinfo{author}{Shone, F.}, \bibinfo{author}{Li, Z.},
  \bibinfo{author}{Frangi, A.F.}, \bibinfo{author}{Xie, S.Q.},
  \bibinfo{author}{Zhang, Z.Q.}, \bibinfo{year}{2023}.
\newblock \bibinfo{title}{Physics-informed deep learning for musculoskeletal
  modeling: Predicting muscle forces and joint kinematics from surface emg}.
\newblock \bibinfo{journal}{IEEE Transactions on Neural Systems and
  Rehabilitation Engineering} \bibinfo{volume}{31}, \bibinfo{pages}{484--493}.
\bibitem[{Zhang et~al.(2018)Zhang, Cui, Pei, Wang and Zhu}]{b30}
\bibinfo{author}{Zhang, Z.}, \bibinfo{author}{Cui, P.}, \bibinfo{author}{Pei,
  J.}, \bibinfo{author}{Wang, X.}, \bibinfo{author}{Zhu, W.},
  \bibinfo{year}{2018}.
\newblock \bibinfo{title}{Timers: Error-bounded svd restart on dynamic
  networks}, in: \bibinfo{booktitle}{Proceedings of the Thirty-Second AAAI
  Conference on Artificial Intelligence and Thirtieth Innovative Applications
  of Artificial Intelligence Conference and Eighth AAAI Symposium on
  Educational Advances in Artificial Intelligence}, \bibinfo{publisher}{AAAI
  Press}.
\bibitem[{Zhou et~al.(2019a)Zhou, Fan, Wang, Wang and
  Li}]{DBLP:journals/www/ZhouFWWL19}
\bibinfo{author}{Zhou, J.}, \bibinfo{author}{Fan, J.}, \bibinfo{author}{Wang,
  J.}, \bibinfo{author}{Wang, X.}, \bibinfo{author}{Li, L.},
  \bibinfo{year}{2019}a.
\newblock \bibinfo{title}{Cost-efficient viral marketing in online social
  networks}.
\newblock \bibinfo{journal}{World Wide Web} \bibinfo{volume}{22},
  \bibinfo{pages}{2355--2378}.
\bibitem[{Zhou et~al.(2019b)Zhou, Liu, Pei et~al.}]{b41}
\bibinfo{author}{Zhou, Y.}, \bibinfo{author}{Liu, W.}, \bibinfo{author}{Pei,
  Y.}, et~al., \bibinfo{year}{2019}b.
\newblock \bibinfo{title}{Dynamic network embedding by semantic evolution}, in:
  \bibinfo{booktitle}{IJCNN}.

\end{thebibliography}



\end{document}